# Optical neuromorphic processing at Tera-OP/s speeds based on Kerr soliton crystal microcombs

Mengxi Tan,[1] Xingyuan Xu,[2] and David J. Moss [1]

[1]Optical Sciences Centre, Swinburne University of Technology, Hawthorn, VIC 3122, Australia
[2]Department of Electrical and Computer Systems Engineering, Monash University, Clayton, 3800 VIC, Australia

## ABSTRACT

Convolutional neural networks (CNNs), inspired by biological visual cortex systems, are a powerful category of artificial neural networks that can extract the hierarchical features of raw data to greatly reduce the network parametric complexity and enhance the predicting accuracy. They are of significant interest for machine learning tasks such as computer vision, speech recognition, playing board games and medical diagnosis [1-7]. Optical neural networks offer the promise of dramatically accelerating computing speed to overcome the inherent bandwidth bottleneck of electronics. Here, we demonstrate a universal optical vector convolutional accelerator operating beyond 10 Tera-OPS (TOPS - operations per second), generating convolutions of images of 250,000 pixels with 8-bit resolution for 10 kernels simultaneously — enough for facial image recognition. We then use the same hardware to sequentially form a deep optical CNN with ten output neurons, achieving successful recognition of full 10 digits with 900 pixel handwritten digit images with 88% accuracy. Our results are based on simultaneously interleaving temporal, wavelength and spatial dimensions enabled by an integrated microcomb source. This approach is scalable and trainable to much more complex networks for demanding applications such as unmanned vehicle and real-time video recognition.

## 1. INTRODUCTION

Artificial neural networks (ANNs) are collections of nodes with weighted connections that, with proper feedback to adjust the network parameters, can "learn" and perform complex operations for face recognition, speech translation, playing board games and medical diagnosis [1-4]. While classic fully connected feedforward networks face challenges in processing extremely high-dimensional data, convolutional neural networks (CNNs), inspired by the (biological) behavior of the visual cortex system, can abstract the representations of input data in their raw form, and then predict their properties with both unprecedented accuracy and greatly reduced parametric complexity [5]. CNNs have been widely applied to computer vision, natural language processing and other areas [6, 7].

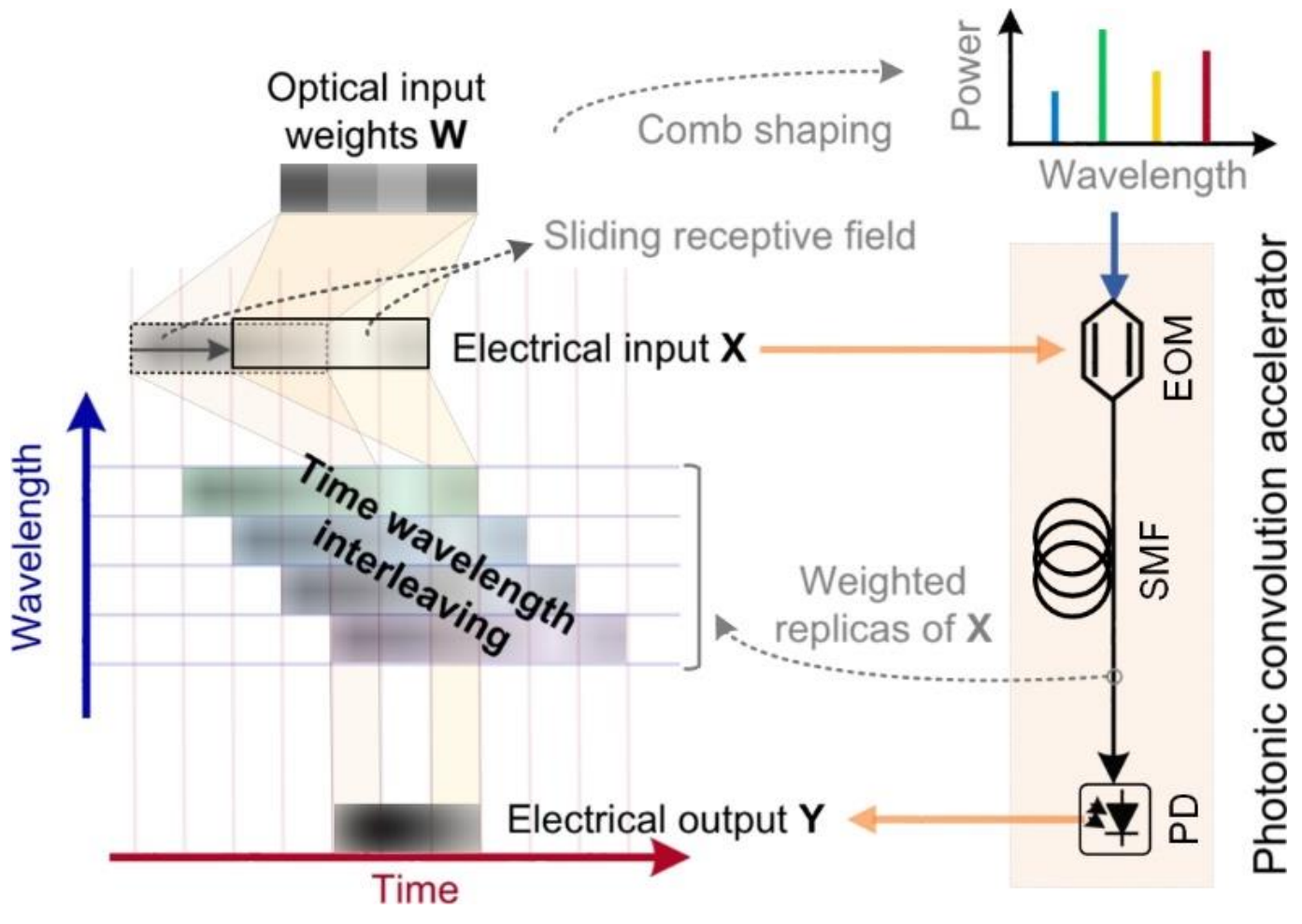

**Figure 1 | Operation principle of the Tera-FLOPS photonic convolution accelerator**. EOM: electro-optical Mach-Zehnder modulator. SMF: standard single mode fibre for telecommunications. PD: photodetector.

The capability of neural networks is dictated by the computing power of the underlying neuromorphic hardware. Optical neural networks (ONNs) [8-18] are promising candidates for next-generation neuromorphic computation, since they have the potential to overcome the bandwidth bottleneck of their electrical counterparts [6, 19-22] and achieve ultra-high computing speeds enabled by the >10 THz wide optical telecom band [8]. ONNs are attracting a great deal of attention with recent breakthroughs and reviews [13-18]. Operating in analog frameworks, they avoid the limitations imposed by the energy and time consumed during reading and storing data back and forth, known as the von Neumann bottleneck [19]. Significant progress has been made in highly parallel, high-speed and trainable ONNs [8-18, 23-27], including approaches that have the potential for full integration on a single photonic chip [8,12,14,15], in turn offering an ultra-high computational density. However, there remains opportunities for significant improvements in ONNs. Processing large-scale data, as needed for practical real-life computer vision tasks, remains challenging because they are primarily fully connected structures where their input scale is determined solely by hardware parallelism. This leads to tradeoffs between the network scale and footprint. Moreover, ONNs have not achieved the extreme computing speeds that analog photonics is capable of.

Here, we demonstrate an optical convolution accelerator to process and compress large-scale data. Through interleaving wavelength, temporal, and spatial dimensions using an integrated Kerr frequency comb, or microcomb [28 – 87], we achieve a vector computing speed as high as 11.322 TOPS and use it to process 250,000 pixel images with 10 convolution kernels at 3.8 TOPs. The convolution accelerator is fully and dynamically reconfigurable, and scalable, and can serve as both a convolutional accelerator front-end with multiple and simultaneous parallel kernels, as well as forming an optically deep CNN with fully connected neurons, with the same hardware. We demonstrate a CNN and successfully apply it to the recognition of full ten digit (0-9) handwritten images, achieving an accuracy of 88%. Our optical neural network represents a major step towards realizing monolithically integrated ONNs and is enabled by our use of an integrated microcomb chip. Moreover, our accelerator scheme is stand alone and universal — fully compatible with either electrical or optical interfaces. Hence, it can serve as a universal ultrahigh bandwidth data compressing front end for any neuromorphic hardware — either optical or electronic — making massive-data machine learning for real-time, ultrahigh bandwidth data possible.

## 2. PRINCIPLE OF OPERATION

Figure 1 shows the operation principle of the photonic convolutional accelerator (CA), featuring high-speed electrical signal input and output data ports, while Figure 2 shows a detailed experimental configuration. The data vector input X is serially encoded with the intensity of temporal symbols in an electrical waveform at a symbol rate $1/\tau$ (baud), where $\tau$ is the symbol period. The convolution kernel is likewise represented by a weight vector W of length R that is used to encode the optical power of the microcomb lines by spectral shaping with a Waveshaper. The temporal waveform X is then multi-cast onto the kernel wavelength channels via electro-optical modulation, generating the replicas weighted by W. Next the optical waveform is transmitted through a dispersive delay with a delay step between adjacent wavelength channels equal to the symbol duration of X, thus achieving time and wavelength interleaving. Finally, the delayed and weighted replicas are summed via high speed photodetection so that each time slot yields a convolution between X and W for a given convolution window, or receptive field. Thus, the convolution window effectively slides at the modulation speed matching the baud rate of X. Each output symbol is the result of R multiply-and-accumulate operations, with the computing speed given by $2R/\tau$ OPS. Since the speed of this process scales with both the baud rate and number of wavelengths, it can be dramatically boosted into the TOP regime by using the massively parallel wavelength channels of a microcomb. Further, the input data X length is unlimited - the convolution accelerator can process arbitrarily large-scale data, limited only by the electronics. Likewise, the kernels number and length are arbitrary, limited only by the number of wavelengths. We achieve simultaneous convolution of multiple kernels by adding additional sub-bands of R wavelengths for each kernel. Following multicasting and dispersive delay, the sub-bands (kernels) are demultiplexed and detected separately with high speed photodetectors, generating a separate electronic waveform for each kernel.

While the convolutional accelerator typically processes vectors, it can operate on matrices for image processing by flattening the matrix into a vector. The precise way that this is done determines both the sliding convolution window's stride and the equivalent matrix computing speed. Our flattening method sets the receptive field (convolution slot) to slide with a horizontal stride of unity (ie., every matrix input element has a corresponding convolution output) and a vertical stride that scales with the size of the convolutional kernel. The larger vertical stride effectively resulted in sub-sampling across the vertical direction of the raw input matrix, equivalent to a partial pooling function [88] in addition to the convolution. This resulted in an effective reduction (or overhead) in matrix computing speed that scales inversely with the size of the kernel, so that a 3x3 kernel results in a speed reduction overhead by 1/3. While this can be eliminated by a variety of means to produce convolutions with a symmetric stride and hence no speed overhead, this is actually not necessary for most applications. Finally, this approach is highly flexible and reconfigurable without any change in hardware - we use same system for the convolutional accelerator for image processing as well as to form an optical deep learning CNN which we use to perform a separate series of experiments. The convolutional accelerator hardware forms both the input processing stage as well as the fully connected neuron layer of the CNN (see below). The system can achieve matrix multiplication by simply sampling one time slot of the output waveform, since the vector dot product is equivalent to the special convolution case where the two input vectors X and W have the same length.

Figure 3 shows a detailed example of the photonic convolution accelerator operating in two different modes. The left panel shows the system performing convolution operations, that are used for the large stand-alone convolution image processing and the convolutional layer of the CNN. The right panel shows the system performing matrix operations which are used as the fully connected layer of the optical CNN. Considering that the experimentally demonstrated configurations are too complex to be presented clearly, in Figure 3 we show a simplified configuration of input data and weights to illustrate the operation principle of our system. The length of **W** and **X** shown in this figure are $R$ = 4 and $L$ = 13 for the case of convolution operations, and $R$ = $L$= 4 for the fully connected layer for matrix operations, respectively.

The schematic of the TOPS photonic convolution accelerator is illustrated in the left panel of Figure 3. The input data vector (length $L$) and weight vector (length $R$) is first multiplexed in the time and wavelength domains, respectively. The input data vector is represented by the intensities of the temporal symbols in a stepwise electrical waveform $\mathbf{X}[n]$ ($n$ denotes discrete temporal locations of the symbols, $n \in [1, L+R-1]$), where $\mathbf{X}[n]$ is the electrical input of the accelerator. The weight vector of the kernel is imprinted onto the optical power of the shaped comb lines as W[$R-i+1$], at the $i^{th}$ wavelength channel ($i \in [1, R]$, where $i$ increases with wavelength).The input electrical waveform $\mathbf{X}[n]$ is first broadcast onto the shaped comb lines via electro-optical modulation. Thus the weighted replica at the $i^{th}$ wavelength channel is W[$R-i+1$]· $\mathbf{X}[n]$. Next, the optical signals across all wavelengths are progressively shifted in the time domain via an optical time-of-flight buffer, which provides a wavelength-sensitive (dispersive) delay with a delay step τ (the difference in delay between adjacent wavelengths) equal to the symbol duration (inverse of the Baud rate) of $\mathbf{X}[n]$. Hence, the shifted replica becomes W[$R-i+1$]· X[$n$-$i$]. Finally, the replicas of all wavelengths are summed via photo-detection as

$$\mathbf{Y}[n] = \sum_{i=1}^{R} \mathbf{W}[R-i+1] \cdot \mathbf{X}[n-i] = (\mathbf{W} * \mathbf{X})[n] \tag{1}$$

where each calculated symbol $\mathbf{Y}[n]$ within the range of [$R$+1, $L$+1] denotes the dot product between **W** and a certain region of **X** (this region is defined by the sliding receptive field as [$n$−R : $n$−1] or [$n$−R, $n$−R+1, $n$−R+2, …, $n$−1]). By simply reading different time slots of the output signal, a convolution is achieved between the weight vector and the input data, thus generating extracted feature maps (matrix convolution outputs) of the input image. While higher order dispersion in the dispersive delay can, in principle, degrade performance, in our experiments this was not a factor.

In addition, the convolution accelerator can also perform matrix multiplication operations, as illustrated in the right panel of Figure 3. The matrix multiplication operations can be treated as a special case of convolution operations when the two input vectors (the pooled and flattened feature maps, and the flattened synaptic weights for the fully connected layer) are the same length ($R$=$L$). Figure 3 shows an example with $R$=$L$=4. Here, we assume the input data vector $\mathbf{X}_{FC}[n]$ and the weight vector $\mathbf{W}_{FC}[R-i+1]$ both have a length $R$ ($i \in [1, R]$, $n \in [1, R]$). Thus, according to Eq. 1, the output waveform after photodetection is

$$\mathbf{Y}_{FC}[n] = \sum_{i=1}^{R} \mathbf{W}_{FC}[R-i+1] \cdot \mathbf{X}_{FC}[n-i] \tag{2}$$

By sampling at the time slot denoted by $n$=$R$+1, the matrix multiplication result of the two input vectors is therefore

$$\mathbf{Y}_{FC}[R+1] = \sum_{i=1}^{R} \mathbf{W}_{FC}[R-i+1] \cdot \mathbf{X}_{FC}[R+1-i] = \sum_{i=1}^{R} \mathbf{W}_{FC}[i] \cdot \mathbf{X}_{FC}[i] \tag{3}$$

## 3. EXPERIMENT

### 3.1 Optical soliton crystal micro-combs

Optical frequency combs, composed of discrete and equally spaced frequency lines, are extremely powerful for optical frequency metrology [28]. Micro-combs offer the full power of optical frequency combs, but in an integrated form with much smaller footprint [28-34]. They have enabled many breakthroughs in high-resolution optical frequency synthesis [32], ultrahigh-capacity communications [33, 34], complex quantum state generation [35 - 43], advanced microwave signal processing [67 - 87], and more. Figure 4 shows a schematic of our optical microcomb chip as well as typical spectra and pumping curves. We use a class of microcomb called soliton crystals that have a crystal-like profile in the angular domain of tightly packed self-localized pulses within micro-ring resonators [34, 47, 48]. They form naturally in micro-cavities with appropriate mode crossings, without complex dynamic pumping or stabilization schemes (described by the Lugiato-Lefever equation [28, 46]). They are characterized by distinctive optical spectra (Fig. 4f) which arise from spectral interference between the tightly packaged solitons circulating along the ring cavity. Soliton crystals exhibit deterministic generation arising from interference between the mode crossing-induced background wave and the high intra-cavity power (Fig. 4c). In turn this enables simple and reliable initiation via adiabatic pump wavelength sweeping [34] that can be achieved with manual detuning (the intracavity power during pump sweeping is shown in Fig. 4d). The key to the ability to adiabatically sweep the pump is that the intra-cavity power is over 30x higher than single-soliton states (DKS), and very close to that of spatiotemporal chaotic states [28, 34]. Thus, the soliton crystal has much less thermal detuning or instability arising from the 'soliton step' that makes resonant pumping of DKS states more challenging. It is this combination of ease of generation and conversion efficiency that makes soliton crystals highly attractive. The coherent soliton crystal microcomb (Figure 4) was generated by optical parametric oscillation in a single integrated MRR (Fig. 4a, 4b) fabricated CMOS-compatible Hydex [22, 23, 34], featuring a Q > 1.5 million, radius 592 µm, and a low FSR of ~ 48.9 GHz. The pump laser (Yenista Tunics – 100S-HP) was boosted by an optical amplifier (Pritel PMFA-37) to initiate the parametric oscillation. The soliton crystal microcomb yielded over 90 channels over the C-band (1540-1570 nm), offering adiabatically generated low-noise frequency comb lines with a small footprint of < 1 $mm^2$ and low power consumption (>100 mW using the technique in [34]).

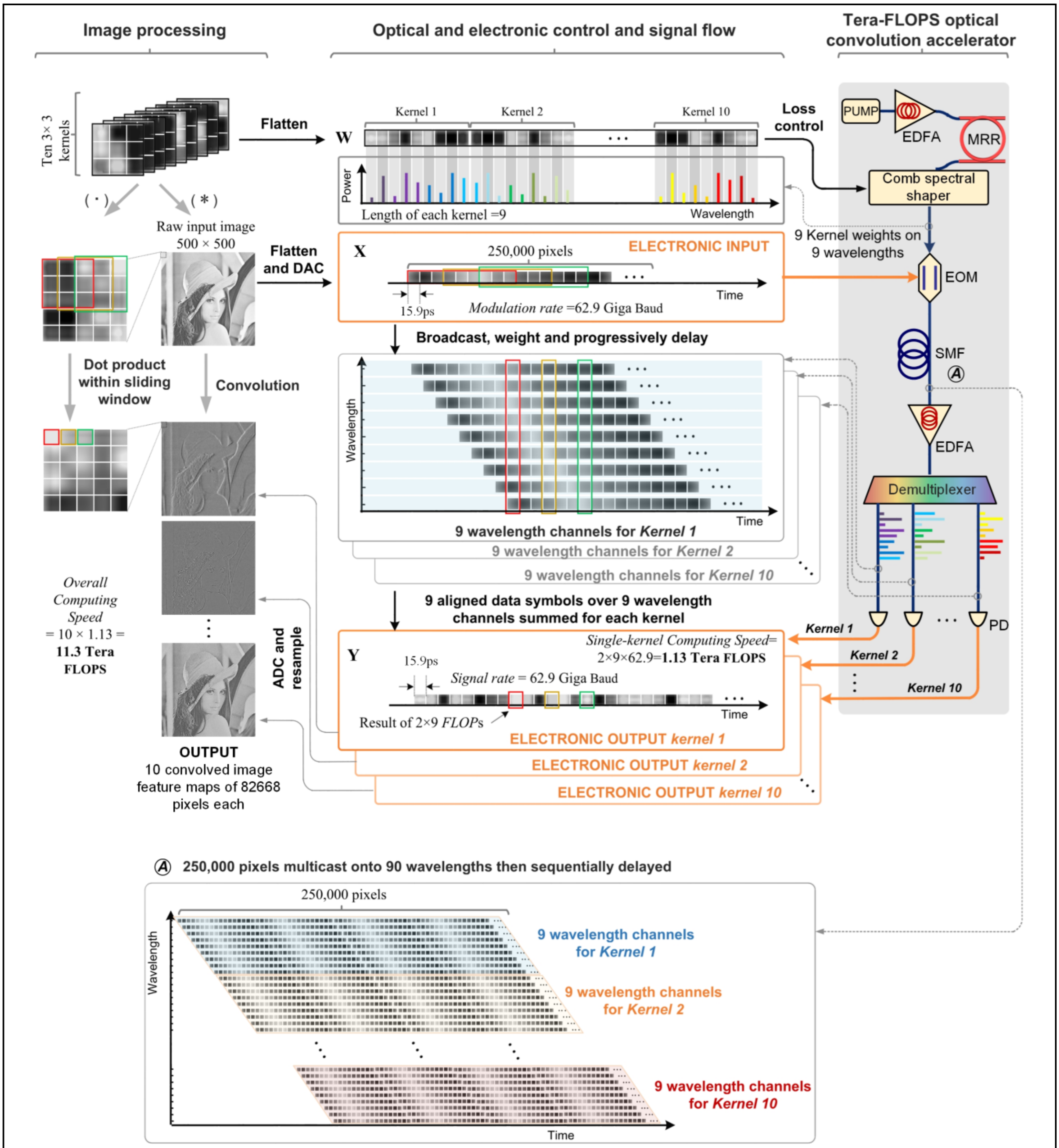

**Figure 2 | Image processing**, consisting of the experimental setup (right panel), the optical and electronic control and signal flow (middle panel), and the corresponding processing flow of the raw input image (left panel). PC: polarization controller. MRR: micro-ring resonator. EOM: electro-optical Mach-Zehnder modulator. SMF: standard single mode fibre.

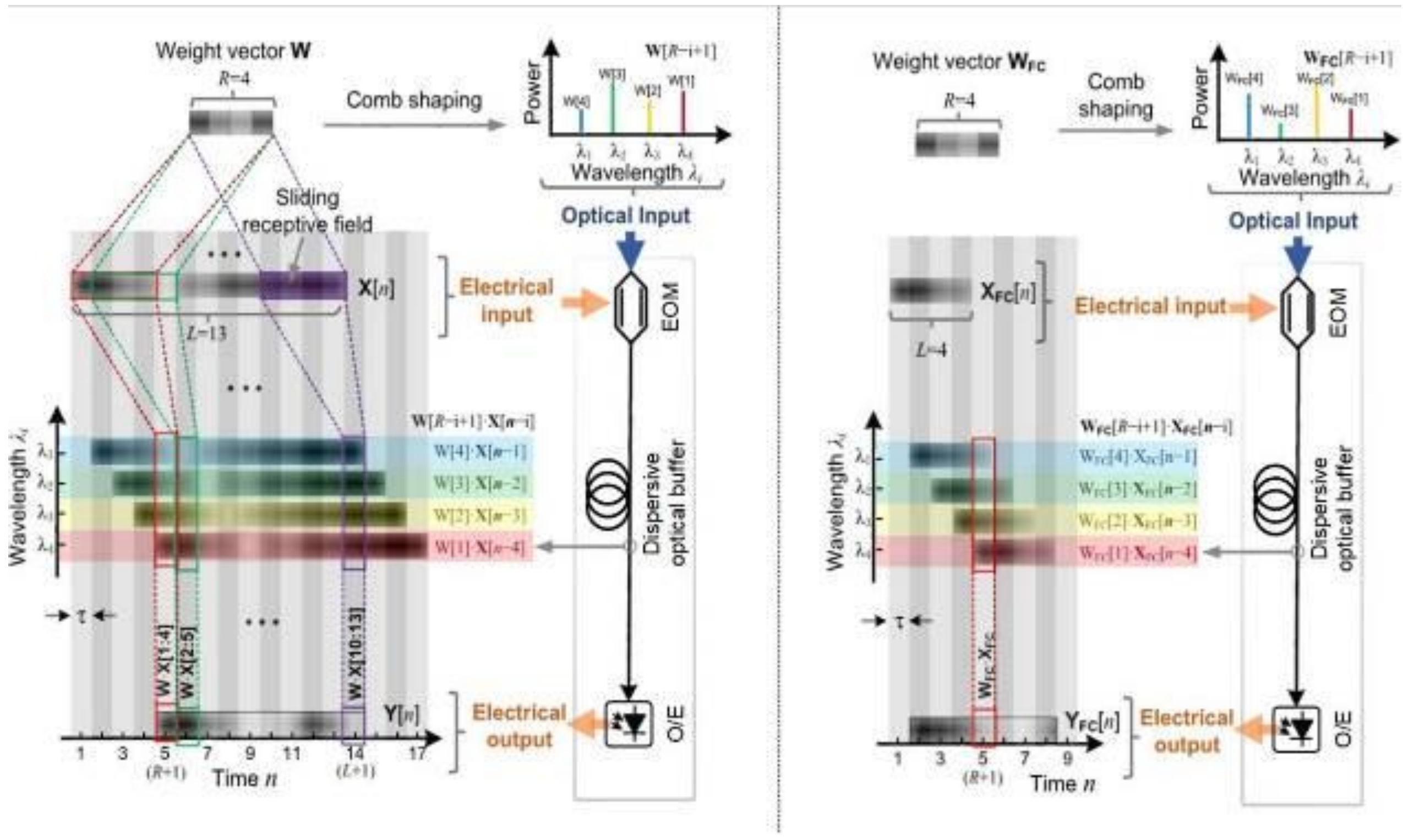

**Figure 3 | TOPS photonic convolution accelerator operating in two modes.** Left: convolution operation mode with $R$ = 4 and $L$ = 13; Right: matrix operation mode with $R$ = $L$ = 4. Electro-optical modulator (EOM), optical buffer that has progressive wavelength-sensitive delay, and an optical-to-electrical conversion module (O/E).

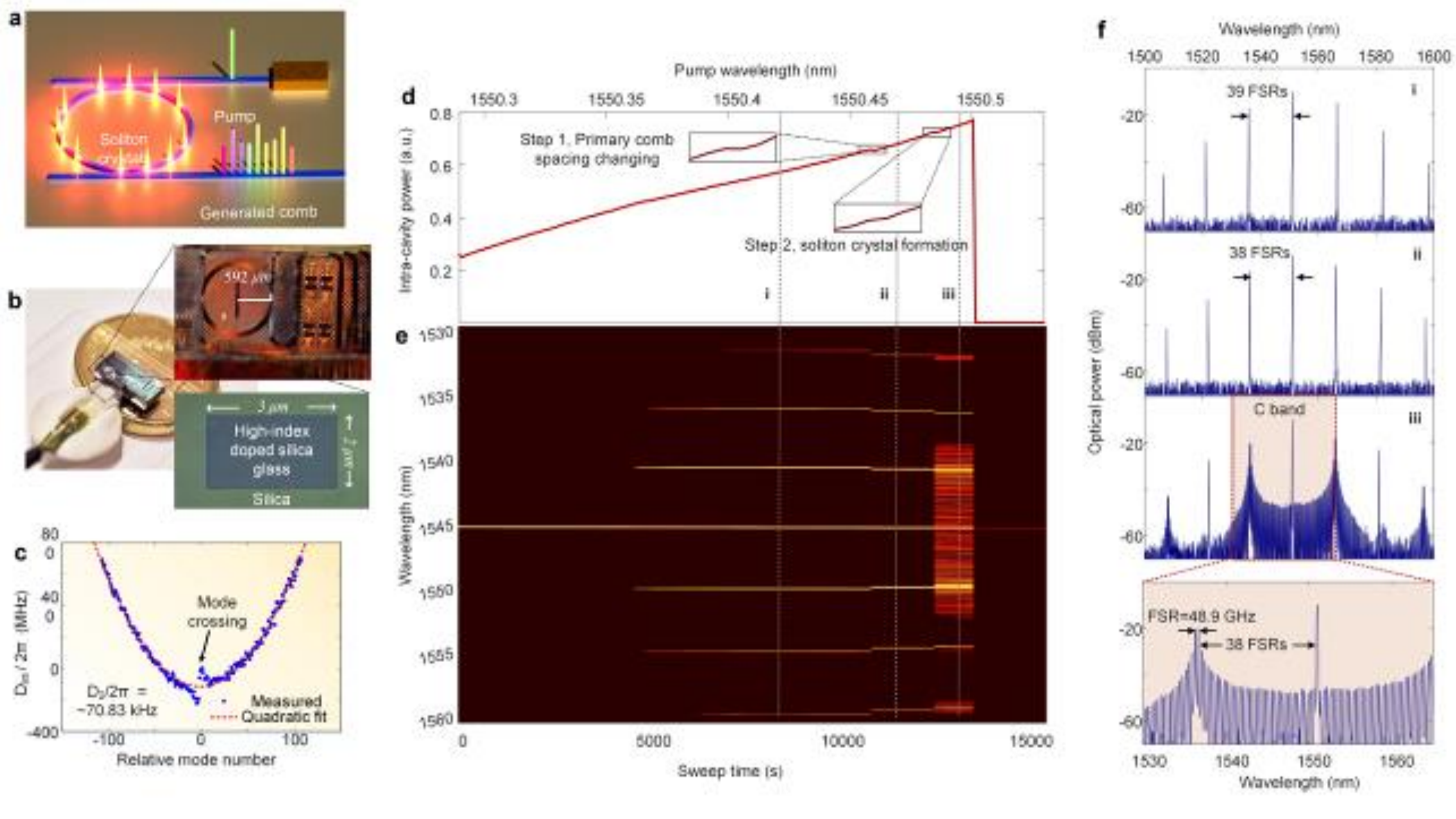

**Figure 4 | a.** Soliton crystal microcomb, generated by pumping an on-chip high-Q nonlinear micro-ring resonator with a CW laser. **b.** Image of the MRR and a SEM image of the MRR cross section. **c.** Measured dispersion of the MRR showing the mode crossing at ~1552 nm. **d.** Measured soliton crystal step of the intra-cavity power, and **e.** optical spectrum of the microcomb when sweeping the pump wavelength. **f.** Optical spectrum of the generated coherent microcomb at different pump detunings at fixed power.

### 3.2 Matrix Convolution Accelerator

Figure 2 shows the experimental setup for the full matrix convolutional accelerator that we use to process a classic 500×500 face image. The system performs 10 simultaneous convolutions with ten 3×3 kernels to achieve distinctive image processing functions. The weight matrices for all kernels were flattened into a composite kernel vector W containing all 90 weights (10 kernels with 3x3=9 weights each), which were then encoded onto the optical power of 90 microcomb lines by an optical spectral shaper (Waveshaper), each kernel occupying its own frequency band of 9 wavelengths. The wavelength channels were supplied by a coherent soliton crystal microcomb (Figure 4) via optical parametric oscillation in a single micro-ring resonator (MRR) (Fig. 4b), radius 592 μm, FSR spacing ~ 48.9 GHz with an optical bandwidth of ~ 36 nm for 90 wavelengths in the C-band (1540-1570 nm) [34].

Figure 5 shows the experimental results of the image processing. Figure 5a depicts the kernel weights and the shaped microcomb's optical spectrum while the input electrical waveform of the image (grey lines are theoretical and blue experimental waveforms) are in Figure 5b. Figure 5c displays the convolved results of the 4$^{th}$ kernel that performs a top Sobel image processing function (grey lines are theory and red experimental). Finally, Figure 5d shows the weight matrices of the kernels and corresponding recovered images.

The raw 500×500 input face image was flattened electronically into a vector X and encoded as the intensities of 250,000 temporal symbols with a resolution of 8 bits/symbol (limited by the electronic arbitrary waveform generator (AWG)), to form the electrical input waveform via a high-speed electrical digital-to-analog converter, at a data rate of 62.9 Giga Baud (time-slot τ =15.9 ps) (Fig. 5b). The waveform duration was 3.975μs for each image corresponding to a processing rate for all ten kernels of over 1/3.975μs, equivalent to 0.25 million of these ultra-large-scale images per second.

The input waveform X was then multi-cast onto the 90 shaped comb lines via electro-optical modulation, yielding replicas weighted by the kernel vector W. Following this, the waveform was transmitted through ~2.2 km of standard single mode fibre having a dispersion of ~17ps/nm/km. The fibre length was carefully chosen to induce a relative temporal shift in the weighted replicas with a progressive delay step of 15.9 ps between adjacent wavelengths, exactly matching the duration of each input data symbol τ, resulting in time and wavelength interleaving for all ten kernels.

The 90 wavelengths were then de-multiplexed into 10 sub-bands of 9 wavelengths, each sub-band corresponding to a kernel, and separately detected by 10 high speed photodetectors. The detection process effectively summed the aligned symbols of the replicas (the electrical output waveform of one of the kernels (kernel 4) is shown in Fig. 5c). The 10 electrical waveforms were converted into digital signals via ADCs and resampled so that each time slot of each of the waveforms corresponded to the dot product between one of the convolutional kernel matrices and the input image within a sliding window (i.e., receptive field). This effectively achieved convolutions between the 10 kernels and the raw input image. The resulting waveforms thus yielded the 10 feature maps (convolutional matrix outputs) containing the extracted hierarchical features of the input image (Figure 5d).

The convolutional vector accelerator made full use of time, wavelength, and spatial multiplexing, where the convolution window effectively slides across the input vector X at a speed equal to the modulation baud-rate — 62.9 Giga Symbols/s. Each output symbol is the result of 9 (the length of each kernel) multiply-and-accumulate operations, thus the core vector computing speed (i.e., throughput) of each kernel is 2×9×62.9 = 1.13 TOPS. For ten kernels computed in parallel the overall computing speed of the vector CA is therefore 1.13×10 =11.3 TOPS, or 11.321×8=90.568 Tb/s (reduced slightly by the optical signal to noise ratio (OSNR)). This speed is over 500 x the fastest ONNs reported to date.

For the image processing matrix application demonstrated here, the convolution window had a vertical sliding stride of 3 (resulting from the 3×3 kernels), and so the effective matrix computing speed was 11.3/3=3.8 TOPs. Homogeneous strides operating at the full vector speed can be readily achieved by duplicating the system with parallel weight-and-delay paths (see below), although we found that this was unnecessary. While the length of the input data processed here was 250,000 pixels, the convolution accelerator can process data with an arbitrarily large scale, the only practical limitation being the capability of the external electronics.

To achieve the designed kernel weights, the generated microcomb was shaped in power using liquid crystal on silicon based spectral shapers (Finisar WaveShaper 4000S). We used two WaveShapers in the experiments - the first was used to flatten the microcomb spectrum while the precise comb power shaping required to imprint the kernel weights was performed by the second, located just before the photo-detection. A feedback loop was employed to improve the accuracy of comb shaping, where the error signal was generated by first measuring the impulse response of the system with a Gaussian pulse input and comparing it with the ideal channel weights. Figure 6 shows the experimental and theoretical large scale facial image processing results achieved by the matrix convolutional accelerator with ten convolutional kernels.

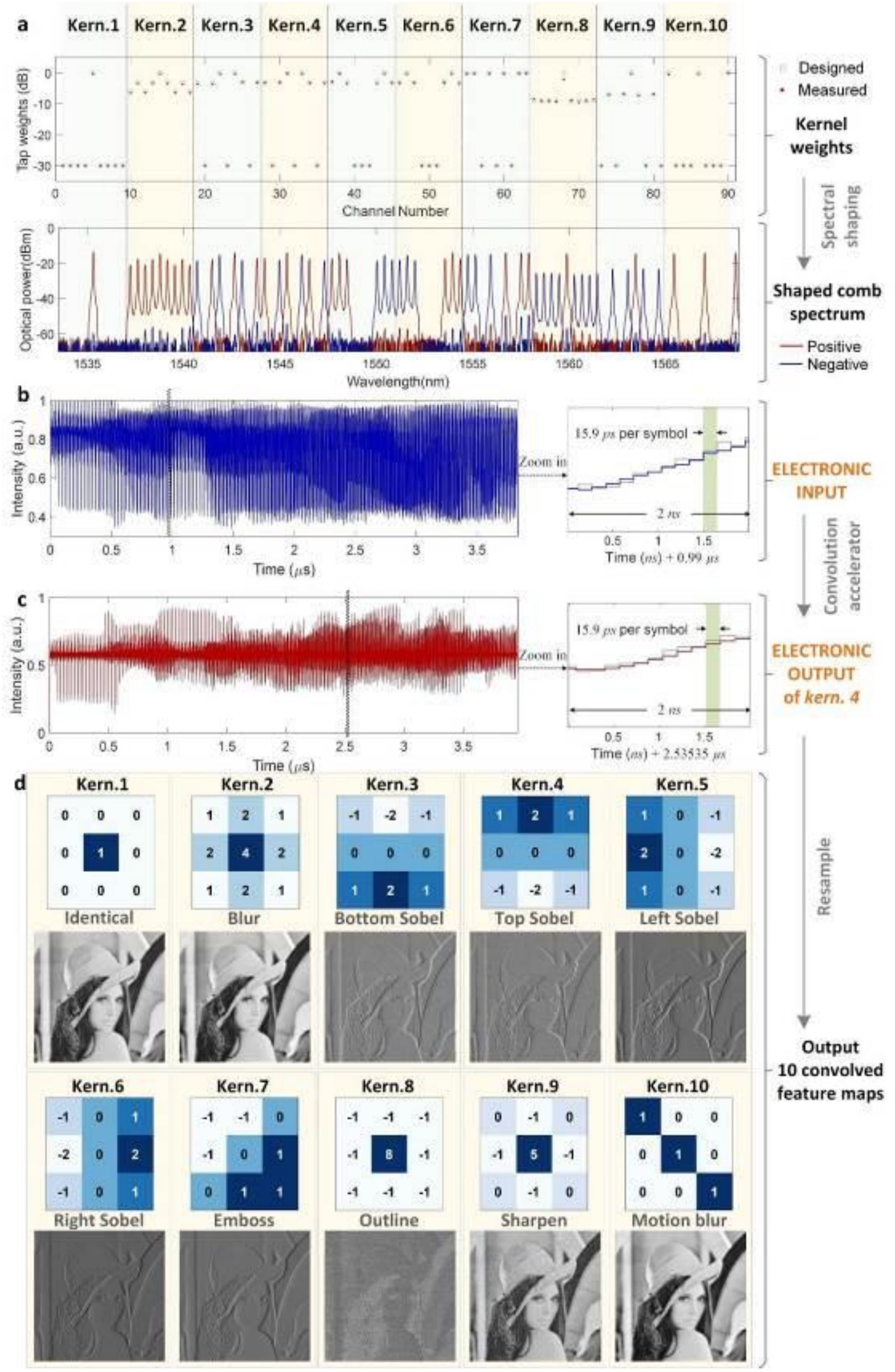

**Figure 5 | Experimental results of the image processing.** a. The kernel weights and the shaped microcomb's optical spectrum. b. The input electrical waveform of the image (grey lines are theory and blue experimental). c. The convolved results of the 4th kernel that performs top Sobel image processing. d. The weight matrices of the kernels and corresponding recovered images.

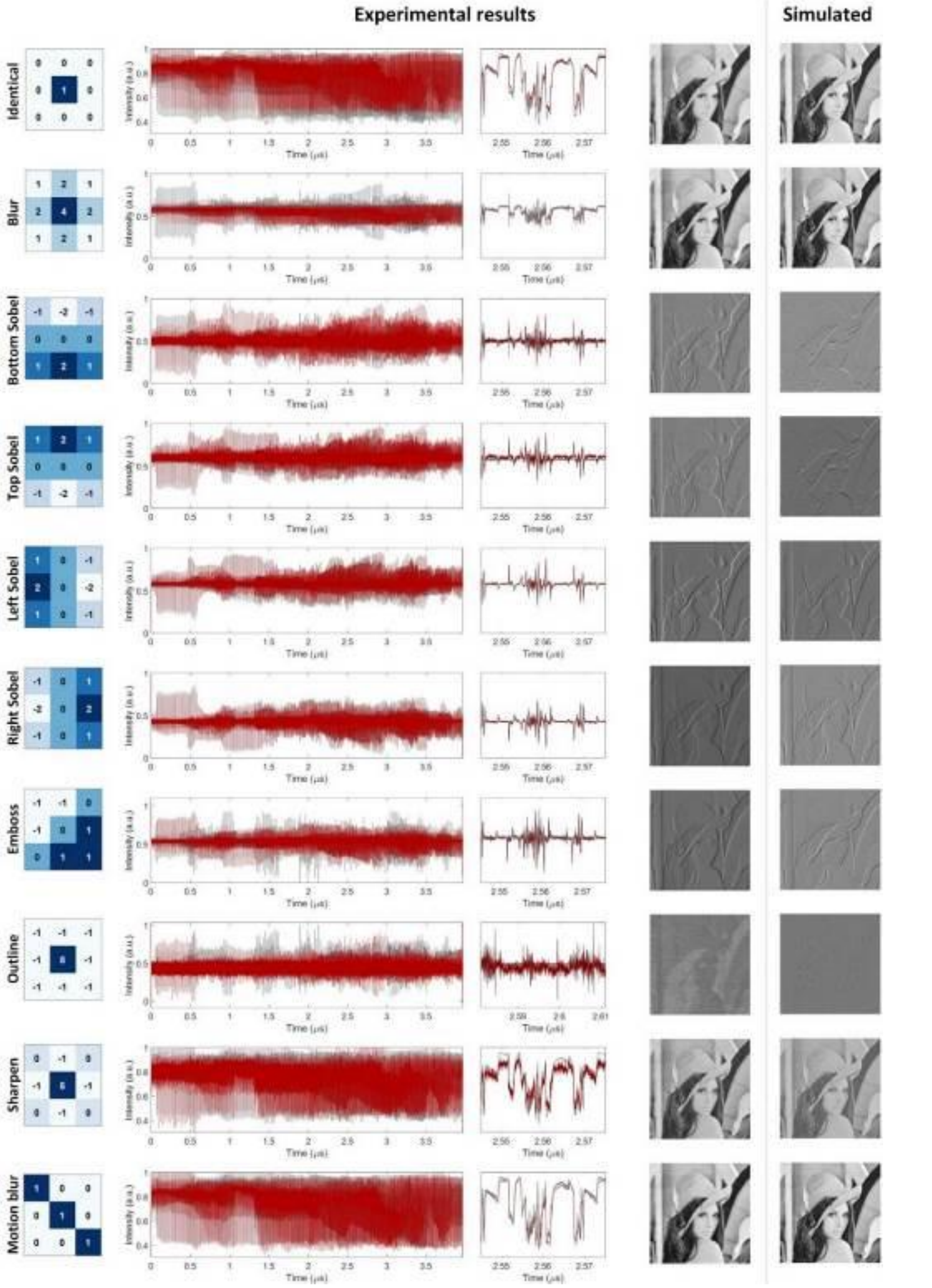

**Figure 6.** Experimental and simulated large scale facial image processing results achieved by the convolutional accelerator with ten convolutional kernels.

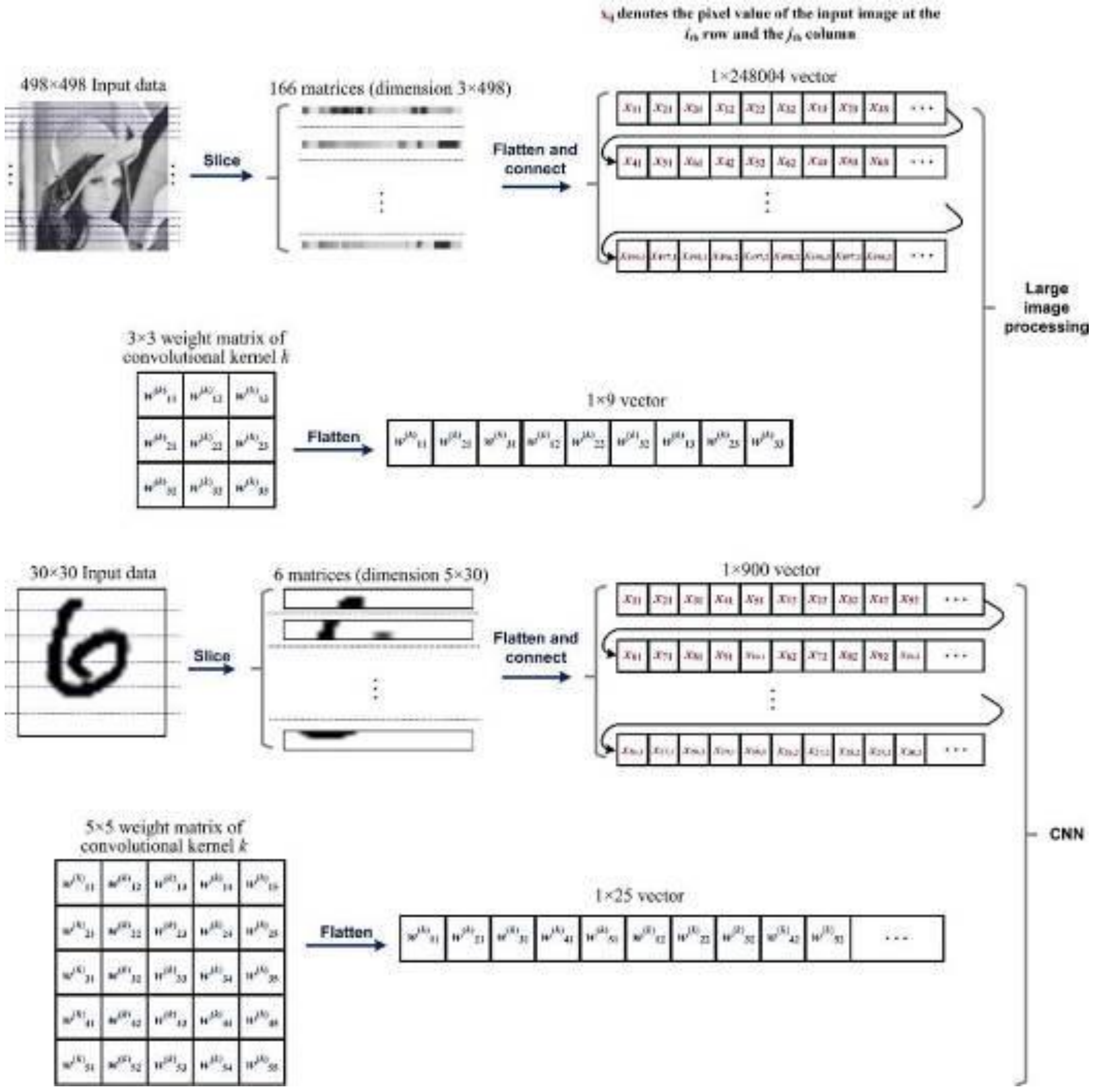

**Figure 7.** Flattening method applied to the input data and kernel matrices. Also see the video presentation.

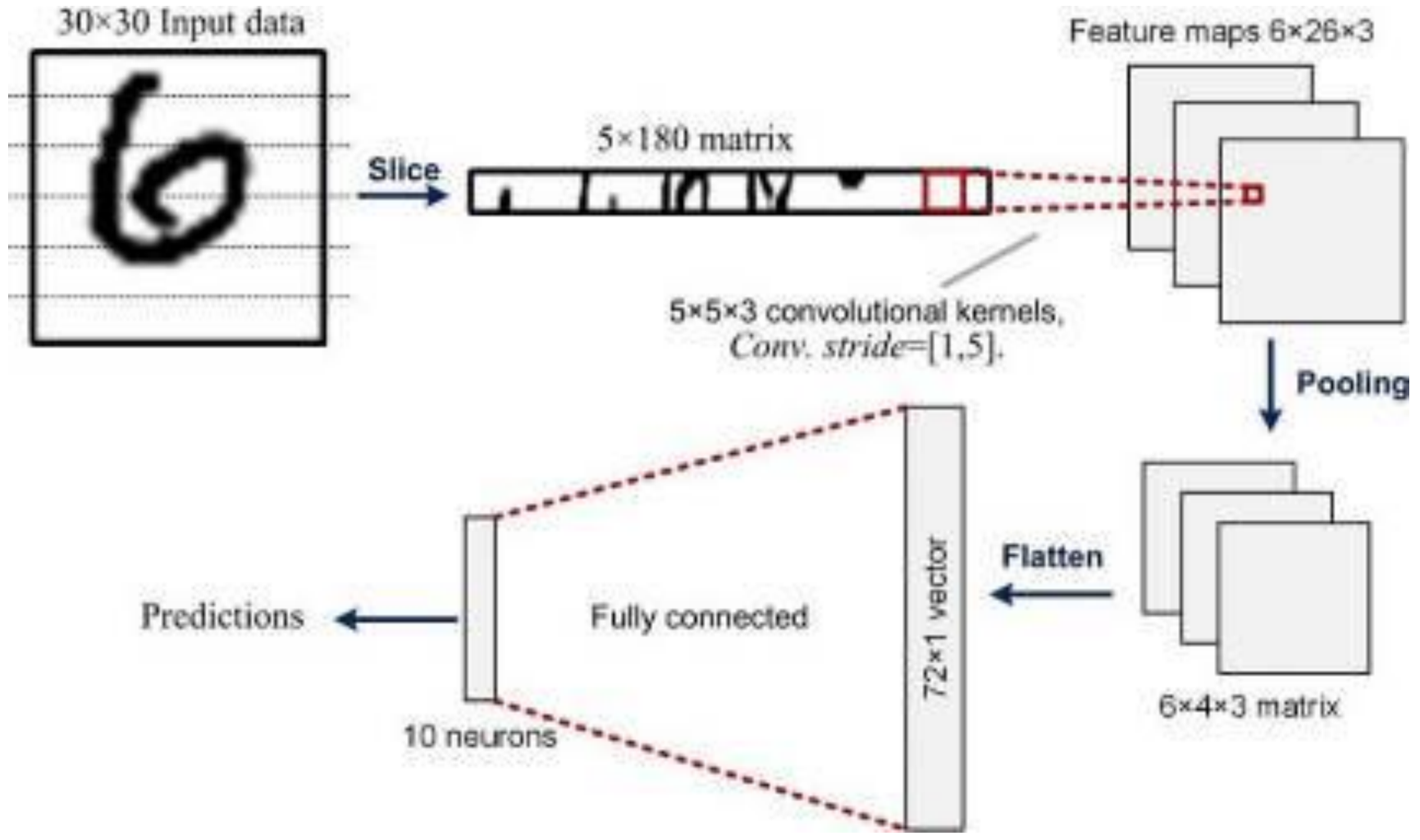

**Figure 8**. Deep (multiple) level CNN structure. The feature maps are the convolutional matrix outputs while the fully connected layers embody the neural network component.

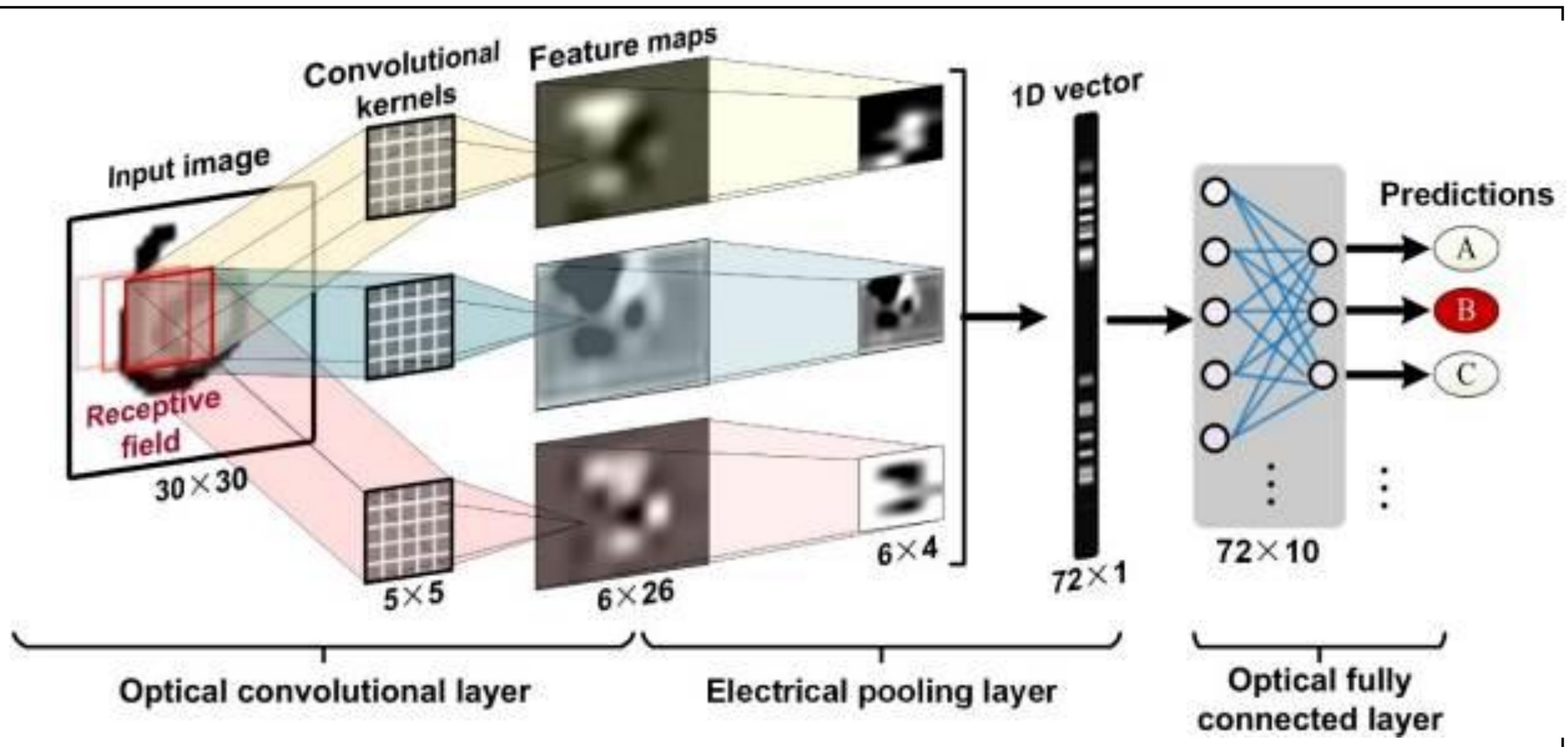

**Figure 9.** The architecture of the optical CNN, including a convolutional layer, a pooling layer, and a fully connected layer.

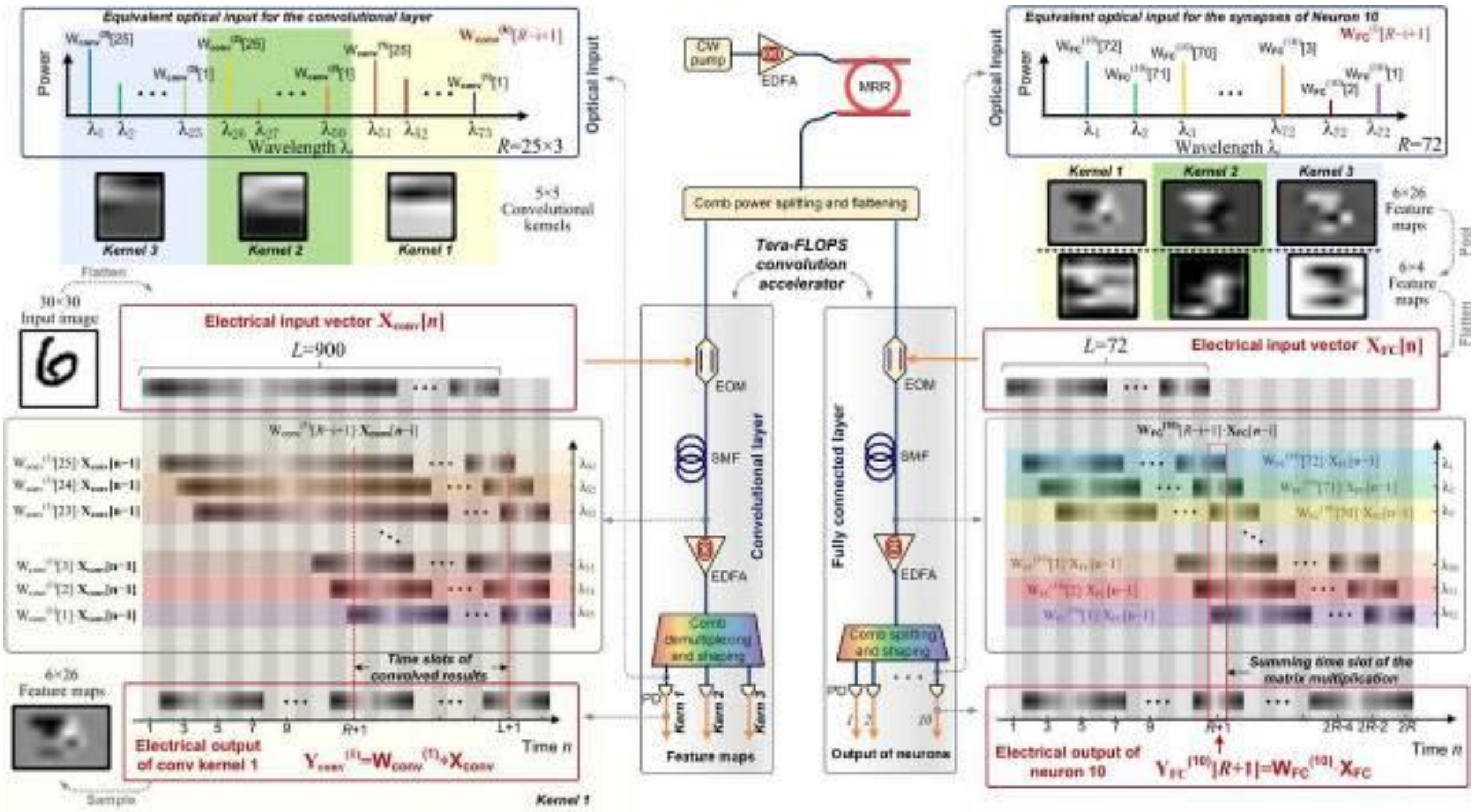

**Figure 10. Experimental schematic of the optical CNN.** Left side is the input front end convolutional accelerator while the right side is the fully connected layer, both of which form the deep learning optical CNN. The microcomb source supplies the wavelengths for both the tera-OPS photonic convolution accelerator as well as the fully connected layer systems. The electronic digital signal processing (DSP) module used for sampling and pooling etc. is external to this structure.

The electrical input data was temporally encoded by an arbitrary waveform generator (Keysight M8195A) and then multicast onto the wavelength channels via a 40 GHz intensity modulator (iXblue). For the 500×500 image processing, we used sample points at a rate of 62.9 Giga samples/s to form the input symbols. We then employed a 2.2 km length of dispersive fibre that provided a progressive delay of 15.9 ps/channel, precisely matched to the input baud rate.

Considering that the convolutional accelerator fundamentally operates on vectors, for applications to image processing, the input data is in the form of matrices and so it needs to be flattened into vectors. (see Figure 7 and also additional video presentation [link]). We follow a common approach where the raw input matrix is first sliced horizontally into multiple sub-matrices, each with a height equal to that of the convolutional kernel. The sub-matrices were then flattened into vectors and connected head-to-tail to form the desired vector (Fig. 7). This flattening method equivalently makes the receptive field slide with a horizontal stride of 1 and a vertical stride equal to the height of the convolutional kernel. We note that a small stride (such as a horizontal stride of 1) ensures that all features of the raw data are extracted, while a large stride (3 or 5) reduces the overlap between the sliding convolution windows and effectively subsamples the convolved feature maps, thus partially serving as a pooling function. A stride of 4 was used for the AlexNet [88].

We note that although the homogeneous strides are generally used more often in digitally implemented CNNs, inhomogeneous convolution strides (unequal horizontal and vertical strides) such as those used here are often used and in most cases, including our experiments, do not limit the performance. Our performance was verified by the high recognition success rate of the CNN for full 10 digit prediction. Further, if desired homogeneous convolutions can be achieved by duplicating the weight-and-delay paths (each including a modulator, a spool of dispersive fibre, a de-multiplexer and multiple photo-detectors) of the accelerator.

### 3.2 Deep Learning Optical Convolutional Neural Network

The convolutional accelerator architecture presented here is fully and dynamically reconfigurable and scalable with the same hardware system. We were thus able to use the accelerator to sequentially form both a frontend convolution processor as well as a fully connected layer, together yielding an optical deep CNN. We applied the CNN to the recognition of full 10 (0-9) handwritten digit images. Figure 8 shows the overall architecture of the deep (multiple) level CNN structure. The feature maps are the convolutional matrix outputs while the fully connected layers embody the neural network component. Figure 9 shows the architecture of the optical CNN, including a convolutional layer, a pooling layer, and a fully connected layer. Figure 10 shows the detailed experimental schematic of the optical CNN. The left side is the input front end convolutional accelerator while the right is the fully connected layer - both the deep learning optical CNN. The microcomb supplies the wavelengths for both the convolution accelerator as well as the fully connected layer. The electronic digital signal processing (DSP) module used for sampling and pooling is external.

The convolutional layer (Fig. 10, left) performs the heaviest computing duty of the entire network, generally taking 55% to 90% of the total computing power. The digit images – 30×30 matrices of grey-scale values with 8-bit resolution – were flattened into vectors and multiplexed in the time-domain at 11.9 Giga Baud (time-slot $\tau$ =84 ps). Three 5×5 kernels were used, requiring 75 microcomb lines, resulting in a vertical stride of 5. The dispersive delay was achieved with ~13 km of SMF to match the data baud-rate. The wavelengths were de-multiplexed into the three kernels which were detected by high speed photodetectors and then sampled and nonlinearly scaled with digital electronics to recover the extracted hierarchical feature maps of the input images. The feature maps were then pooled electronically and flattened into a vector (Eq. 2,3) $X_{FC}$ (72×1= 6×4×3) per image that formed the input data to the fully connected layer.

The fully connected layer had 10 neurons, each corresponding to one of the 10 categories of handwritten digits from 0 to 9, with the synaptic weights represented by a 72×10 weight matrix $\mathbf{W}_{FC}^{(l)}$ (ie., ten 72×1 column vectors) for the $l^{th}$ neuron ($l \in [1, 10]$) – with the number of comb lines (72) matching the length of the flattened feature map vector $\mathbf{X}_{FC}$. The shaped optical spectrum at the $l^{th}$ port had an optical power distribution proportional to the weight vector $\mathbf{W}_{FC}^{(l)}$, thus serving as the equivalent optical input of the $l^{th}$ neuron. After being multicast onto the 72 wavelengths and progressively delayed, the optical signal was weighted and demultiplexed with a single Waveshaper into 10 spatial output ports — each corresponding to a neuron. Since this part of the network involved linear processing, the kernel wavelength weighting could be implemented either before the EO modulation or at a later stage just before photodetection. The advantage of the latter is that both the demultiplexing and weighting can then be achieved with a single Waveshaper. Finally, the different node/neuron outputs were obtained by sampling the 73$^{rd}$ symbol of the convolved results. The final output of the optical CNN was represented by the intensities of the output neurons (Figure 11), where the highest intensity for each tested image corresponded to the predicted category. The peripheral systems, including signal sampling, nonlinear function and pooling, were implemented electronically with digital signal processing hardware, although some of these functions (e.g., pooling) can be performed in the optical domain with the VCA. Supervised network training was performed offline electronically (see below).

We experimentally tested 50 x 8-bit resolution images each 30 × 30 of the handwritten digit dataset with the deep optical CNN. The confusion matrix (Figure 12) shows an accuracy of 88% for the generated predictions, in contrast to 90% for the numerical results calculated on an electrical digital computer. The computing speed of the CA component of the deep optical CNN was 2×75×11.9 =1.785 TOPS, or 14.3 Tb/s. To process image matrices with 5×5 kernels, the convolutional layer had a matrix flattening overhead of 5, yielding an image computing speed of 1.785/5= 357 Giga OPS. The computing speed of the fully connected layer was 119.8 Giga-OPS (see below). The waveform duration was 30×30×84ps=75.6ns for each image, and so the convolutional layer processed images at the rate of 1/75.6ns = 13.2 million handwritten digit images per second.

We note that handwritten digit recognition, although widely employed as a benchmark test in digital hardware, is still (for full 10 digit (0 - 9) recognition) beyond the capability of existing analog reconfigurable ONNs. Digit recognition requires a large number of physical parallel paths for fully-connected networks (e.g., a hidden layer with 10 neurons requires 9000 physical paths), which poses a huge challenge for current nanofabrication techniques. Our CNN represents the first reconfigurable and integrable ONN capable not only of performing high level complex tasks such as full handwritten digit recognition, but at TOP speeds.

For the convolutional layer of the CNN, we used 5 sample points at 59.421642 Giga Samples/s to form each single symbol of the input waveform, which also matched with the progressive time delay (84 ps) of the 13km dispersive fibre. The generated electronic waveforms for 50 images are shown as Fig. 16 and 17, which served as the electrical input signal for the convolutional and fully connected layers, respectively.

For the convolutional accelerator in both the CA and CNN experiments - the 500×500 image processing experiment and the convolutional layer of the CNN - the second Waveshaper simultaneously shaped and de-multiplexed the wavelength channels into separate spatial ports according to the configuration of the convolutional kernels. As for the fully connected layer, the second Waveshaper simultaneously performed the shaping and power splitting (instead of de-multiplexing) for the ten output neurons. The de-multiplexed or power-split spatial ports were sequentially detected and measured. However, these two functions could readily be achieved in parallel with a commercially available 20-port optical spectral shaper (WaveShaper 16000S, Finisar) and multiple photodetectors.

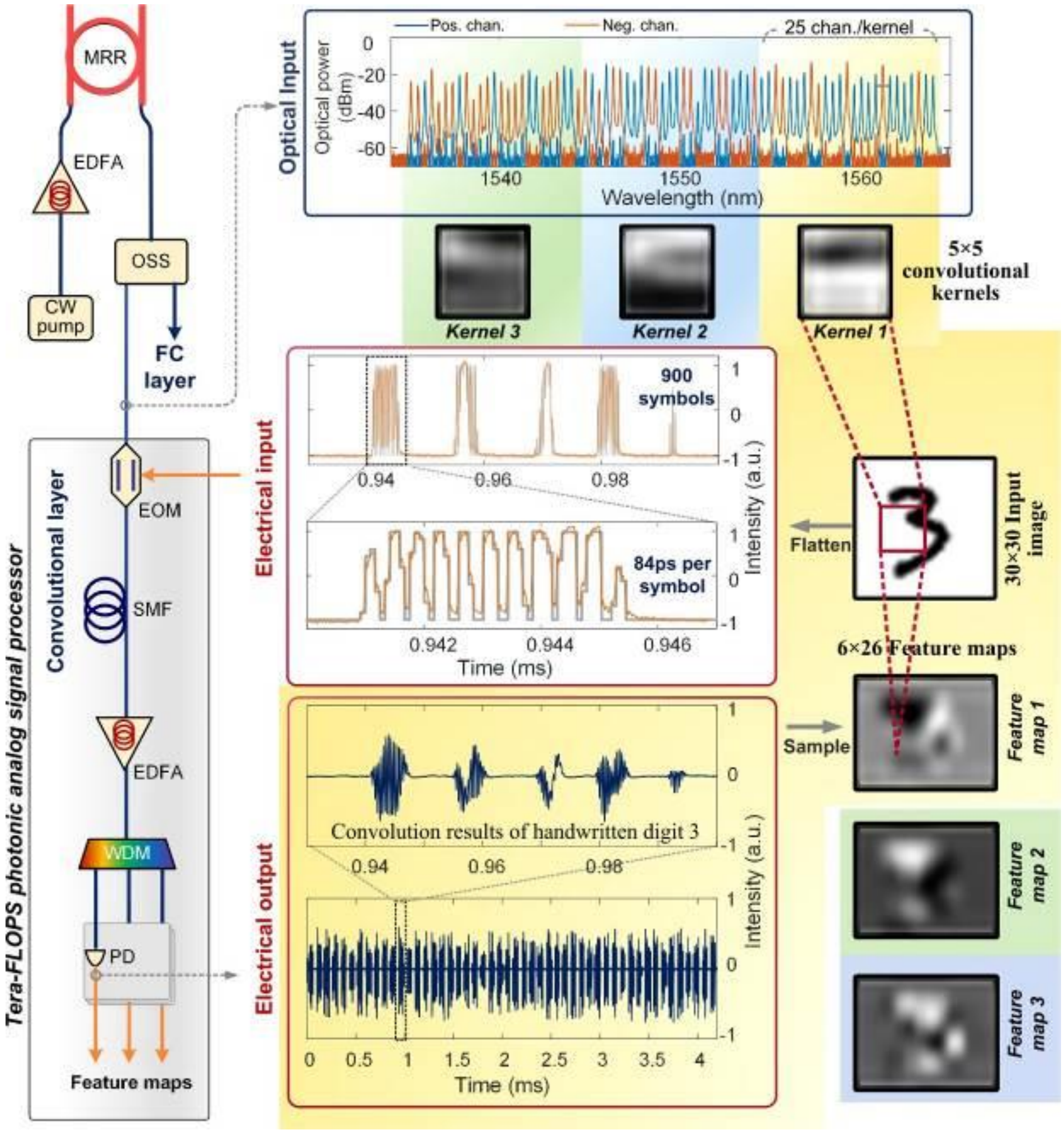

**Figure 11. Convolutional layer. Architecture and experimental results.** Left panel: experimental setup. Right panel: experimental results of one convolutional kernel, showing shaped microcomb optical spectrum and corresponding kernel weights (blue lines are positive, red negative synaptic weights), the input electrical waveform for the digit “3” (middle: grey lines theory, yellow experiment), convolved results and feature maps.

The negative channel weights were achieved using two methods. For the 500×500 image processing experiment and the convolutional layer of the CNN, the wavelength channels of each kernel were separated into two spatial outputs by the WaveShaper according to the signs of the kernel weights, and then detected by a balanced photodetector (Finisar XPDV2020). Conversely, for the fully connected layer the weights were encoded in the symbols of the input electrical waveform during the electrical digital processing stage. Both of these methods to impart negative weights were successful. Finally, the electrical output waveform was sampled and digitized by a high-speed oscilloscope (Keysight DSOZ504A, 80 Giga Symbols/s) to extract the final convolved output. For the CNN, the extracted outputs of the convolution accelerator were further processed digitally, including rescaling to exclude the loss of the photonic link via a reference bit, and then mapped onto a certain range using a nonlinear tanh function. The pooling layer's functions were also implemented digitally, following the algorithm introduced in the network model. The residual discrepancy between experiment and calculations, for both the recognition and convolving functions, was due to the deterioration of the input waveform caused by performance limitations of the electrical arbitrary waveform generator. Addressing this would lead to greater accuracy and closer agreement with numerical calculations.

**Network training and digital processing**

For the deep learning (multiple level) optical CNN, we employed datasets from the MNIST (Modified National Institute of Standards and Technology) handwritten digit database [89]. The dataset contained 60000 images as the training set and 10000 images as the test set. The structure of the CNN in this work (Figure 8) was determined empirically using trial-and-error, which is a standard approach for neural networks. In our case this was greatly aided by the fact that the network structure (number of synapses and neurons) can be reconfigured dynamically without any change in hardware. The 28×28 input data was first padded with zeros into a 30×30 image and then sliced into a 5×180 matrix and convolved with the 5×5 kernels. This slicing operation equivalently made the receptive field slide horizontally with a stride = 1 across the rows and a vertical stride = 5 across the columns of the 30×30 input data (corresponding to the 900 input nodes). Then the 6×26×3 feature map was pooled (using average pooling) to a smaller dimension of 6×4×3. Finally, the matrix was further flattened into a 72×1 vector that served as input nodes for the fully connected layer, which in turn generated the predictions using the 10 output neurons. The nonlinear function we used after the convolutional layer, the pooling function and the fully connected layer was the tanh function. Although other nonlinear functions such as ReLU are widely used, we used this tanh function since it can be realized with a saturating electrical amplifier.

The training necessary to acquire pre-trained weights and biases was performed offline with a digital computer. The Back Propagation algorithm [90] was employed to adjust the weights. To validate the hyper-parameters of the CNN, we performed a 10-fold cross validation using the 60000 samples of the training dataset, where the training set was separated into 10 subsets and each was then used to test the trained network (6000 samples) with the rest of the 9 subsets (54000 samples). The result is shown in Figure 13. The test sets were assessed by both the optical CNN (50 images) and an electronic computer (10000 images, Figure 12) for comparison.

Figure 6 shows the experimental and simulated large scale facial image processing results achieved by the convolutional accelerator with ten convolutional kernels. It shows the experimental results of large 500×500 face image processing, including the recorded waveforms and the recovered images. Figs. 16-21 show the full experimental results of the CNN. Figure 16 shows the shaped impulse response of the convolutional layer that has 3 kernels and 75 wavelengths, or weights, in total. Figure 17 shows the shaped impulse responses for the ten neurons, each with 72 synapses, at the fully connected layer. The fifty handwritten digits tested during our experiments are shown in Fig. 18 (a), with their corresponding encoded electrical waveform shown in Fig. 18 (b), which served as the electrical input of the convolutional layer. The electronic waveform generated from the extracted feature maps is shown in Fig. 19, that served as the input of the fully connected layer. Figure 20 shows the full experimental results of the CNN including the recorded waveforms and the recovered/sampled outputs. Figure 21 shows the experimental results of the ten output neurons in the fully connected layer. The left graphs show the output waveforms, while the right graphs show the corresponding sampled (red dot) and theoretically calculated (grey dot) intensities of the neurons.

Since there are no common standards in the literature for classifying and quantifying the computing speed and processing power of ONNs, we explicitly outline the performance definitions that we use in characterizing our performance. We follow the approach that is widely used to evaluate electronic micro-processors. The computing power of the convolution accelerator—closely related to the operation bandwidth—is denoted as the throughput, which is the number of operations performed within a certain period. Considering that in our system the input data and weight vectors originate from different paths and are interleaved in different dimensions (time, wavelength, and space), we use the temporal sequence at the electrical output port to define the throughput in a more straightforward manner.

At the electrical output port, the output waveform has L+R−1 symbols in total (L and R are the lengths of the input data vector and the kernel weight vector, respectively), among which L−R+1 symbols are the convolution results. Further, each output symbol is the calculated outcome of R multiply-and-accumulate operations or 2R OPS, with a symbol duration τ given by that of the input waveform symbols. Thus, considering that L is generally much larger than R in practical convolutional neural networks, the term (L−R+1)/(L+R−1) would not affect the vector computing speed, or throughput, which (in OPS) is given by

$$\frac{2R}{\tau}\cdot\frac{L-R+1}{L+R-1}\approx\frac{2R}{\tau} \tag{4}$$

As such, the computing speed of the vector convolutional accelerator demonstrated here is 2×9×62.9×10 = 11.321 Tera-OPS for ten parallel convolutional kernels).

We note that when processing data in the form of vectors, such as audio speech, the effective computing speed of the accelerator would be the same as the vector computing speed 2R/ τ. Yet when processing data in the form of matrices, such as for images, we must account for the overhead on the effective computing speed brought about by the matrix-to-vector flattening process. The overhead is directly related to the width of the convolutional kernels, for example, with 3-by-3 kernels, the effective computing speed would be ~1/3 * 2R/τ, which still is in the TOP regime due to the high parallelism brought about by the time-wavelength interleaving technique.

For the convolutional accelerator, the output waveform of each kernel (with a length of L−R+1=250,000−9+1=249,992) contains 166×498=82,668 useful symbols that are sampled out to form the feature map, while the rest of the symbols are discarded. As such, the effective matrix convolution speed for the experimentally performed task is slower than the vector computing speed of the convolution accelerator by the overhead factor of 3, and so the net speed then becomes 11.321×82,668/249,991=11.321×33.07% = 3.7437 TOPS.

For the deep CNN the convolutional accelerator front end layer has a vector computing speed of 2×25×11.9×3 = 1.785 TOPS while the matrix convolution speed for 5x5 kernels is 1.785×6×26/(900−25+1) = 317.9 Giga-OPS. For the fully connected layer of the deep CNN, according to Eq. (4), the output waveform of each neuron would have a length of 2R−1, while the useful (relevant output) symbol would be the one locating at R+1, which is also the result of 2R operations. As such, the computing speed of the fully connected layer would be 2R / (τ*(2R−1)) per neuron. With R =72 during the experiment and ten neurons simultaneous operating, the effective computing speed of the matrix multiplication would be 2R / (τ*(2R−1)) × 10 = 2×72 / (84ps* (2×72−1)) = 119.83 Giga-OPS.

In addition, the intensity resolution (bit-resolution for digital systems) for analog ONNs is mainly limited by the signal-to-noise ratio (SNR). To achieve 8-bit resolution, the SNR of the system needs to be > 20·log10(28) = 48 dB. This was achieved by our accelerator and so our speed in Tb/s is close to the speed in OPs x 8 – not reduced by our OSNR.

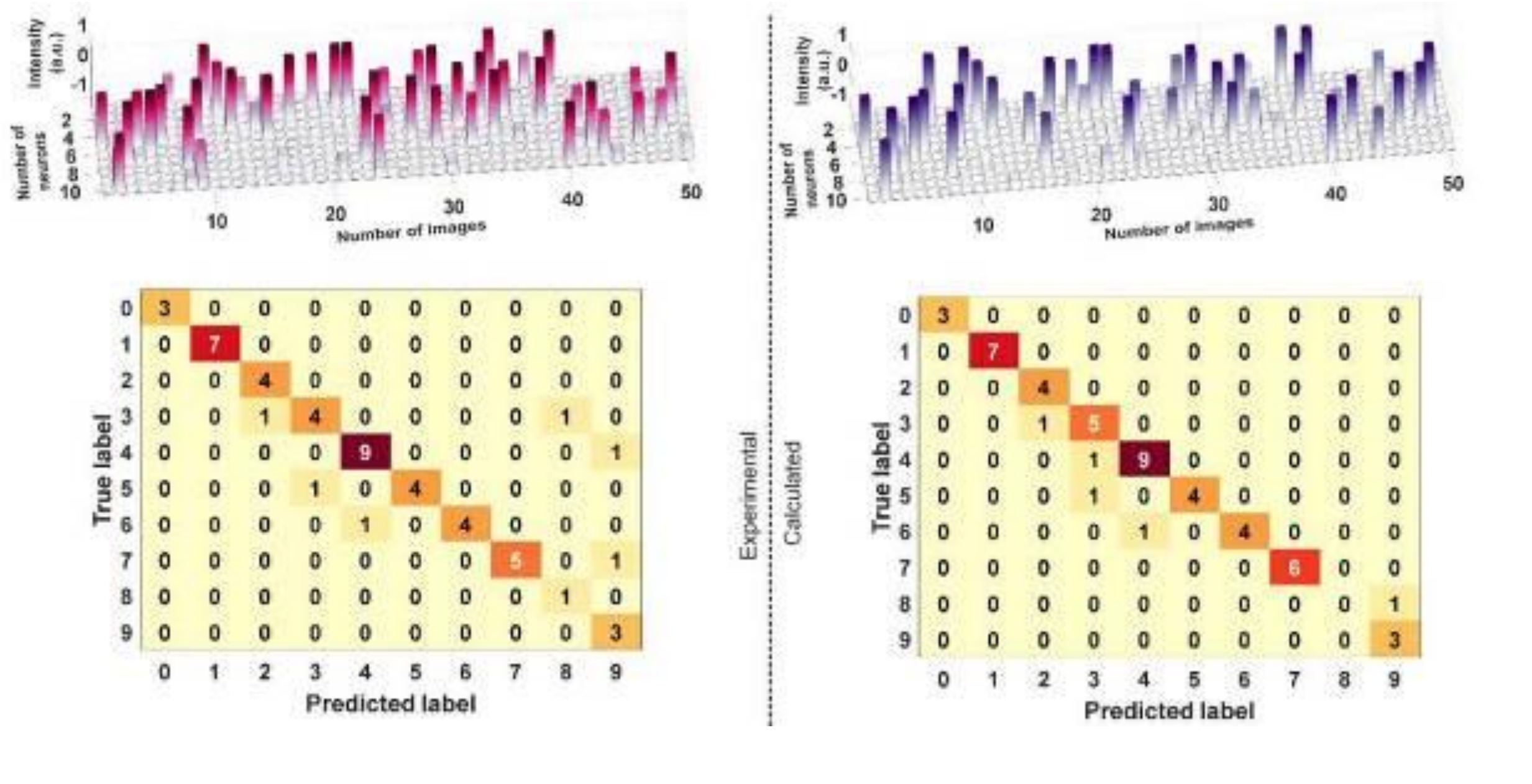

**Figure 12** Experimental and theoretical results for image recognition. The upper figures show the sampled intensities of the ten output neurons at the fully connected layer, while the lower figures show the confusion matrices with the darker colours indicating a higher recognition score.

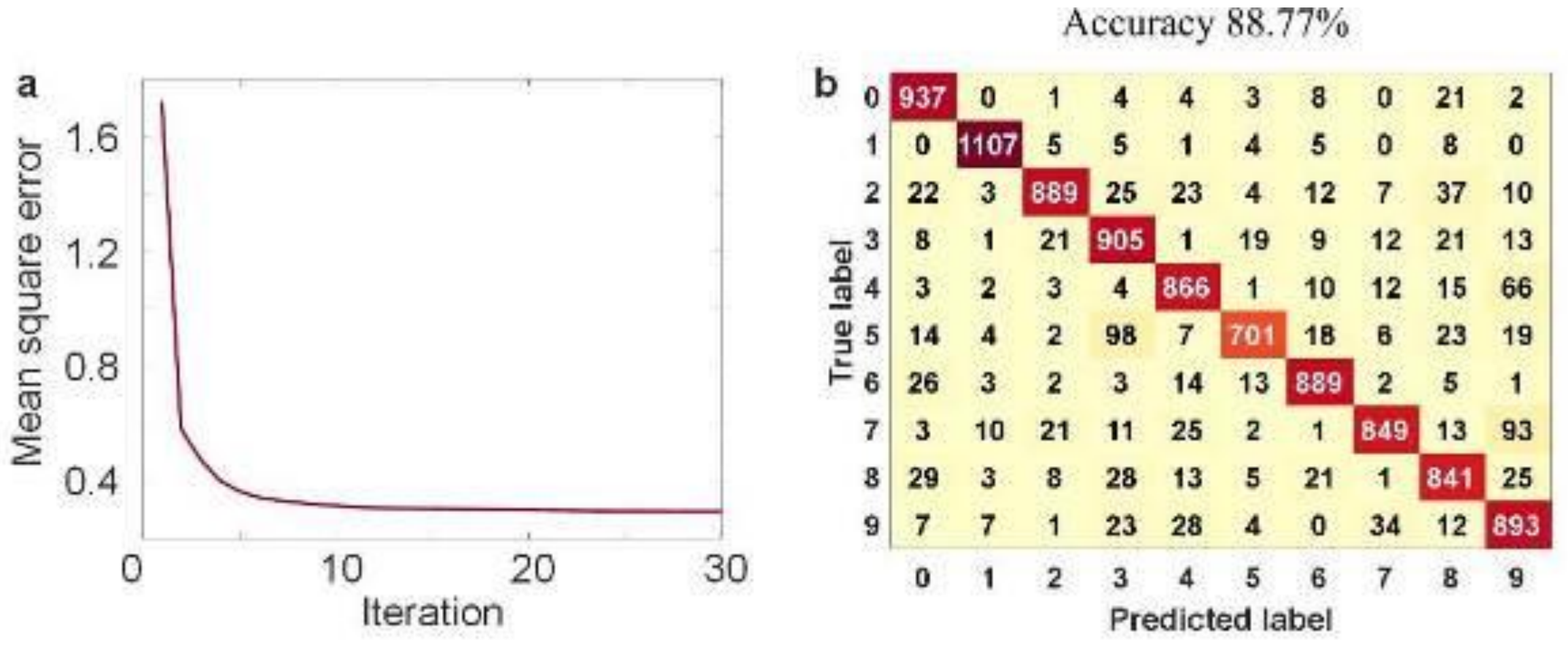

**Figure 13 Results of full 10 digit (0-9) recognition by the deep level optical CNN.** (a) The offline training curve and (b) generated predictions (confusion matrix) obtained by using an electronic computer to process the output of the CNN. The numbers represent the number of predicted labels versus the actual label (digit), hence the diagonal elements represent correct predictions.

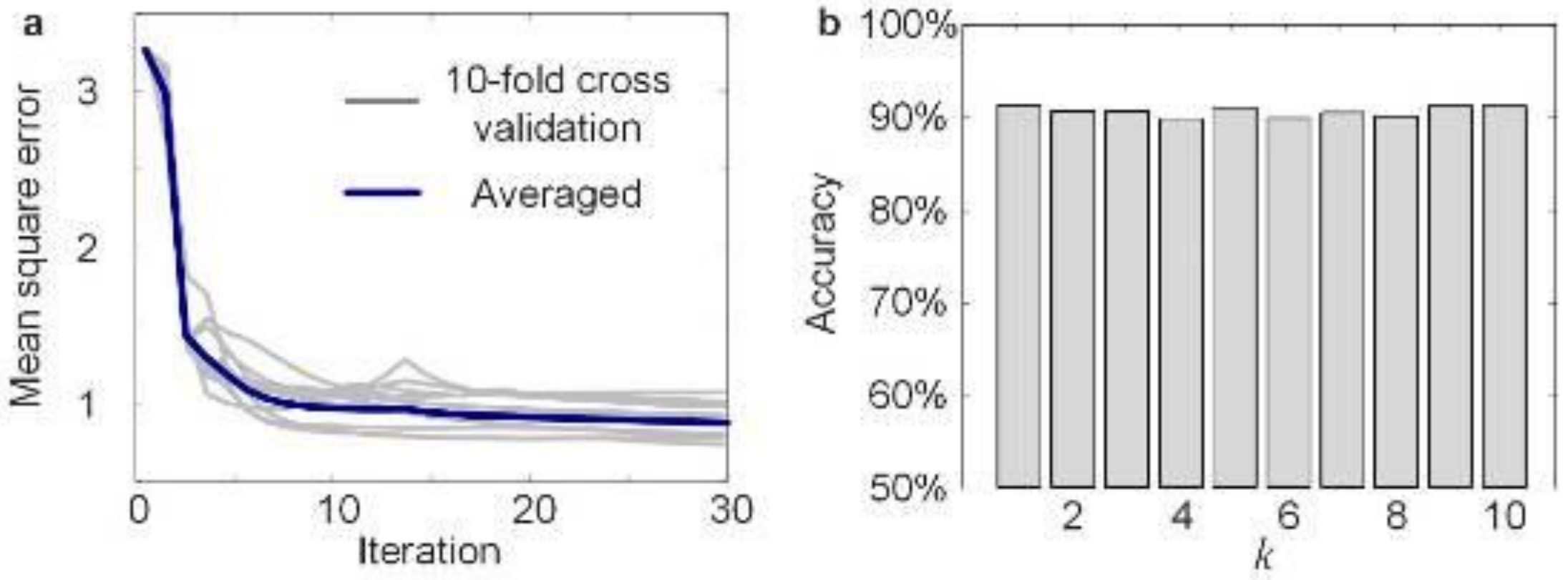

**Figure 14.** (a) Training curve and (b) the accuracy, of using the $k_{th}$ subset as the test set during the 10-fold cross validation for the full 10digit image recognition by the deep CNN.

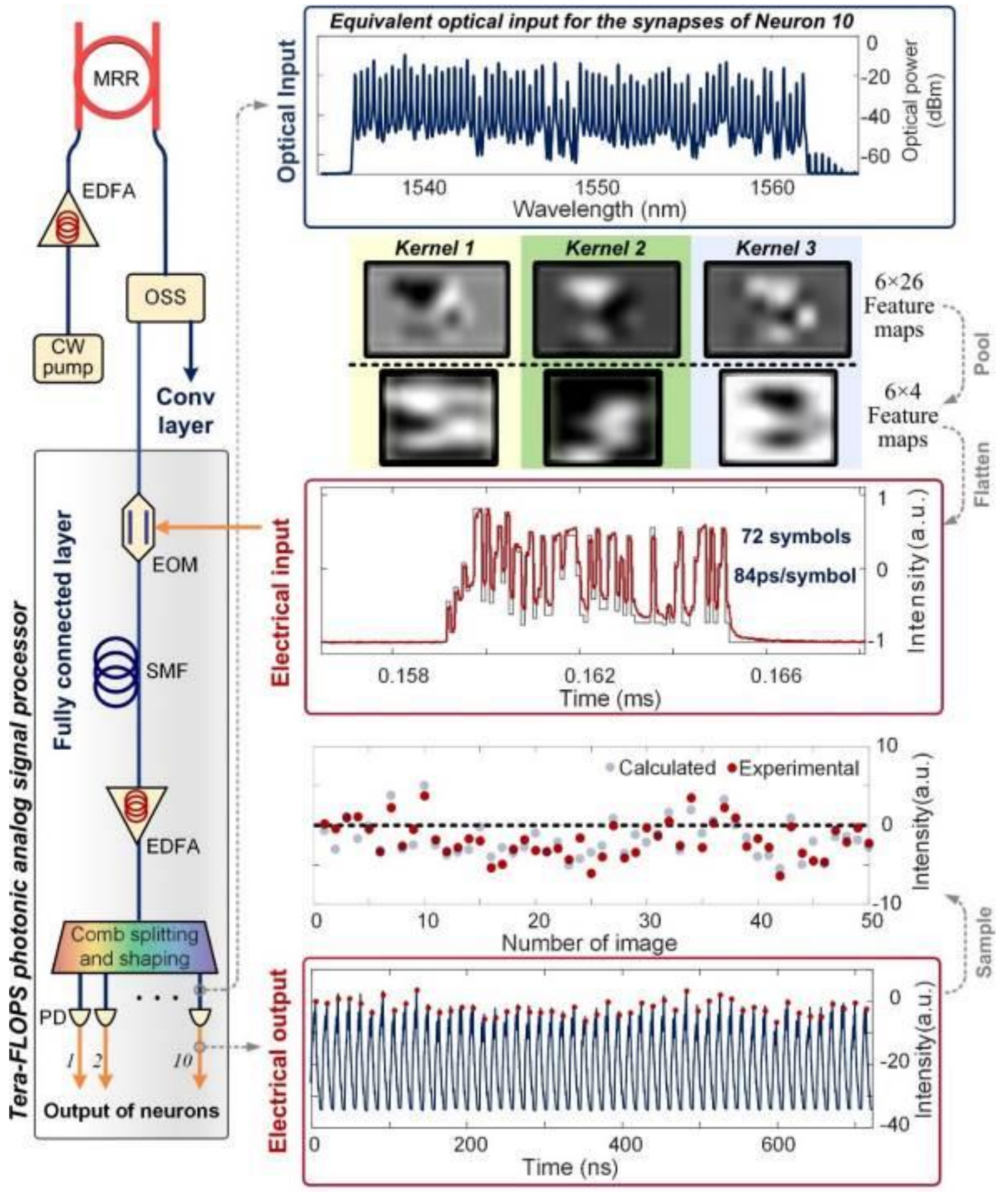

**Figure 15. Fully connected layers. Architecture and experimental results.** The left panel depicts the experimental setup, similar to the convolutional layer. The right panel shows top: the experimental results for one output neuron, including the shaped comb spectrum; middle: the pooled feature maps of the digit “3” and the corresponding input electrical waveform (the grey lines are theory and red experimental; bottom: output waveform of the neuron and sampled intensities.

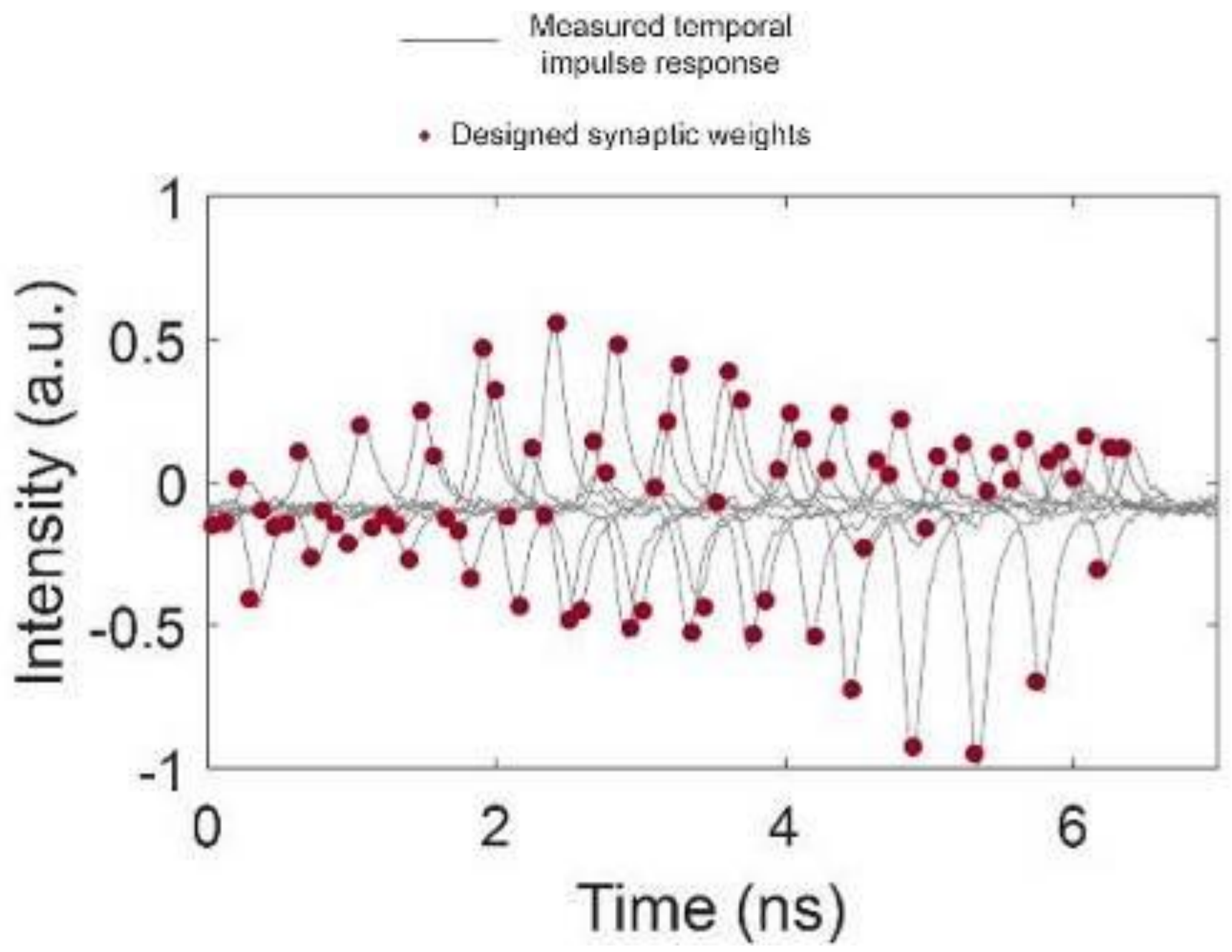

**Fig. 16** Convolutional layer shaped impulse response.

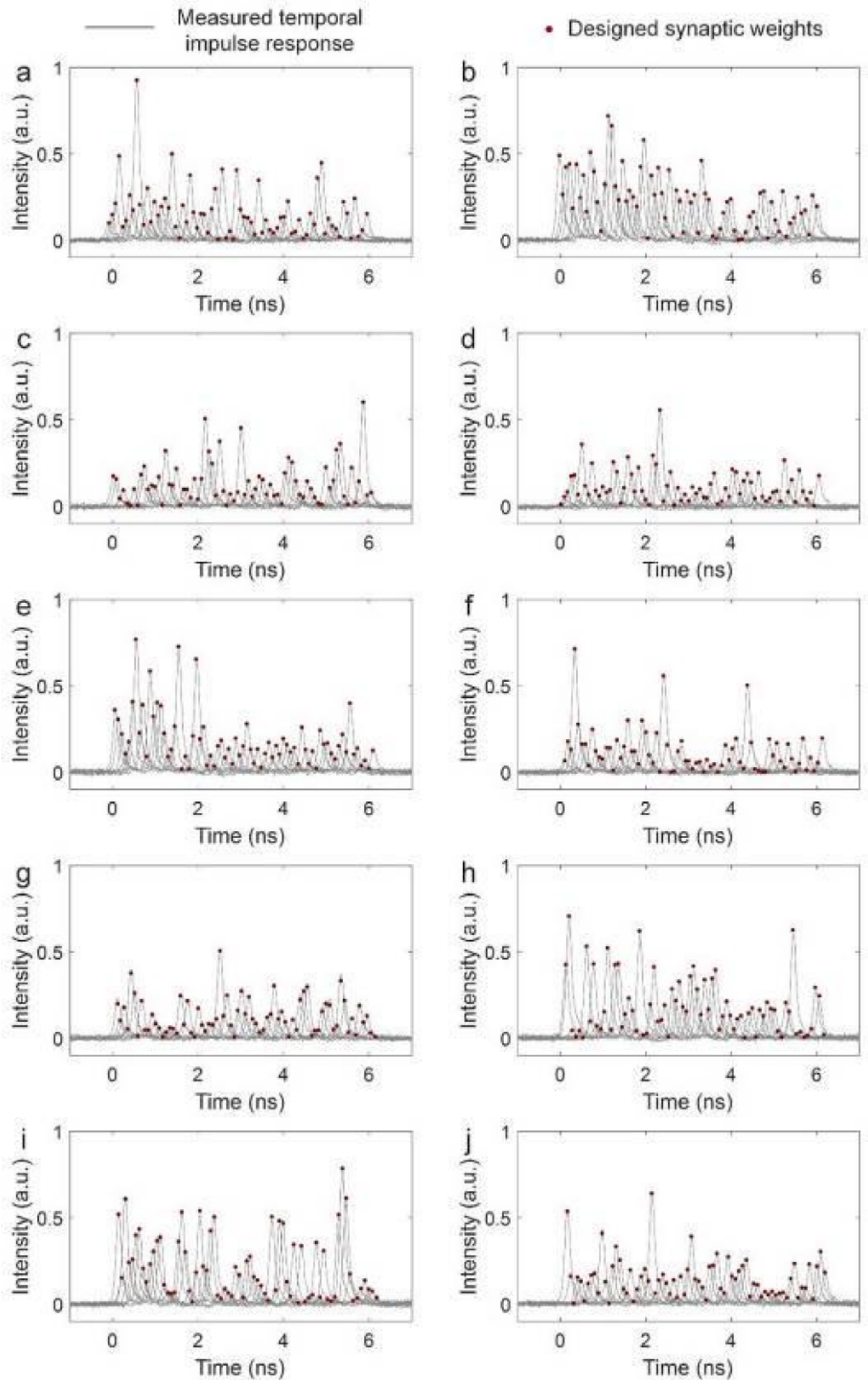

**Fig. 17.** The shaped impulse response of the ten neurons at the fully connected layer.

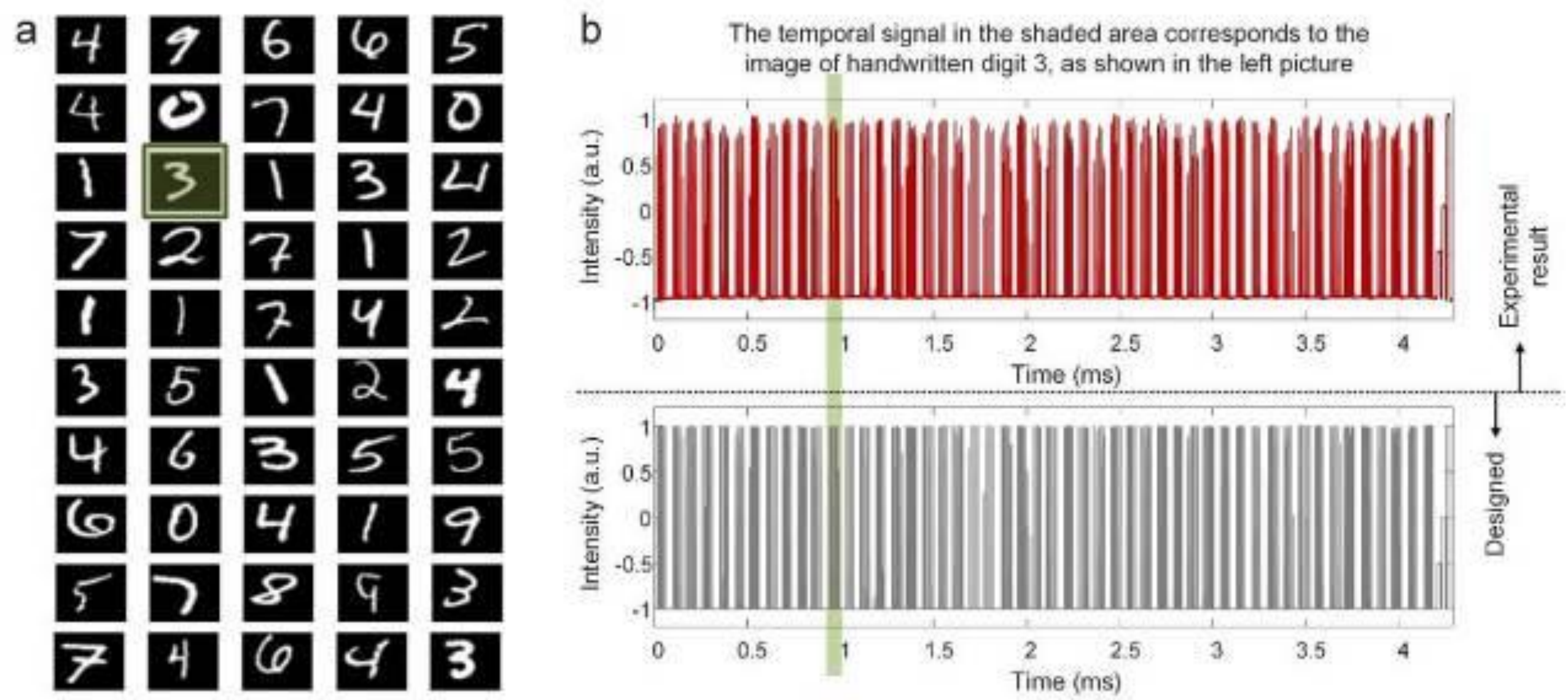

**Fig. 18.** (a) Experimentally recognized 50 handwritten digit images (b) generated waveform that served as the input of the convolutional layer.

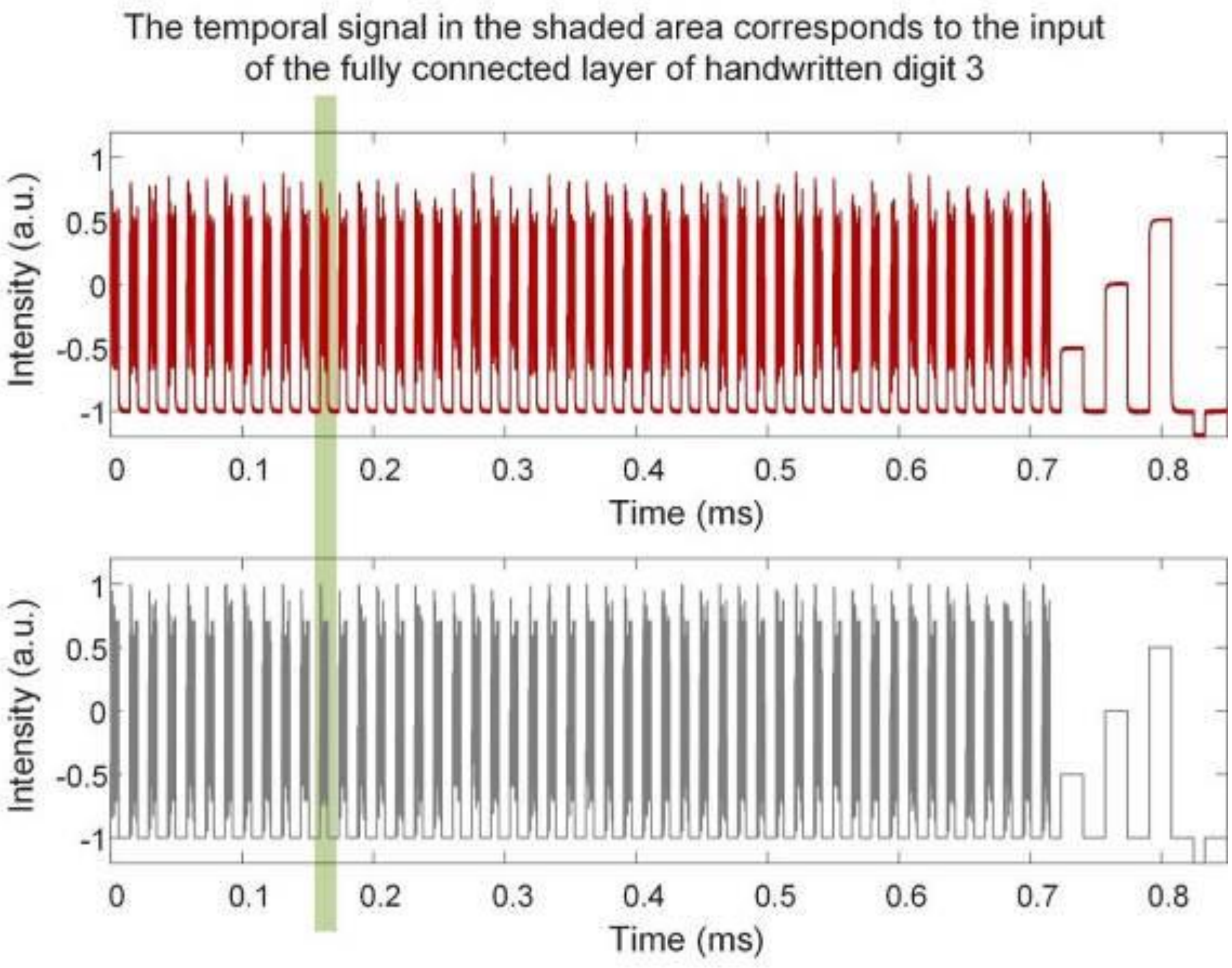

**Fig. 19.** Generated (red) and simulated (grey) electrical waveform that served as the input for the 10$^{th}$ neuron in the fully connected layer.

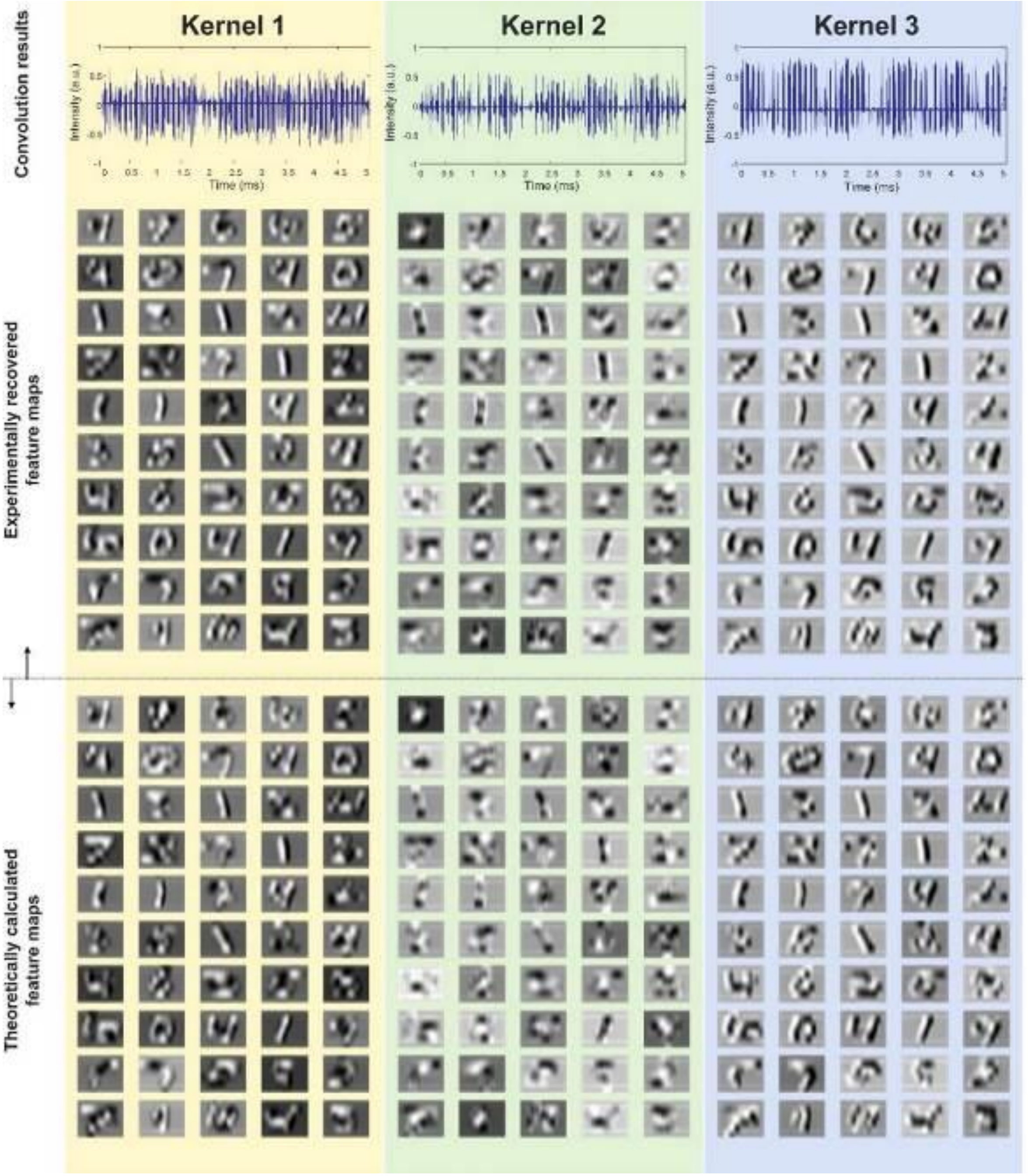

**Fig. 20.** Experimental results of the convolutional layer, including the output waveforms of the convolved results and the corresponding recovered and theoretically calculated feature maps.

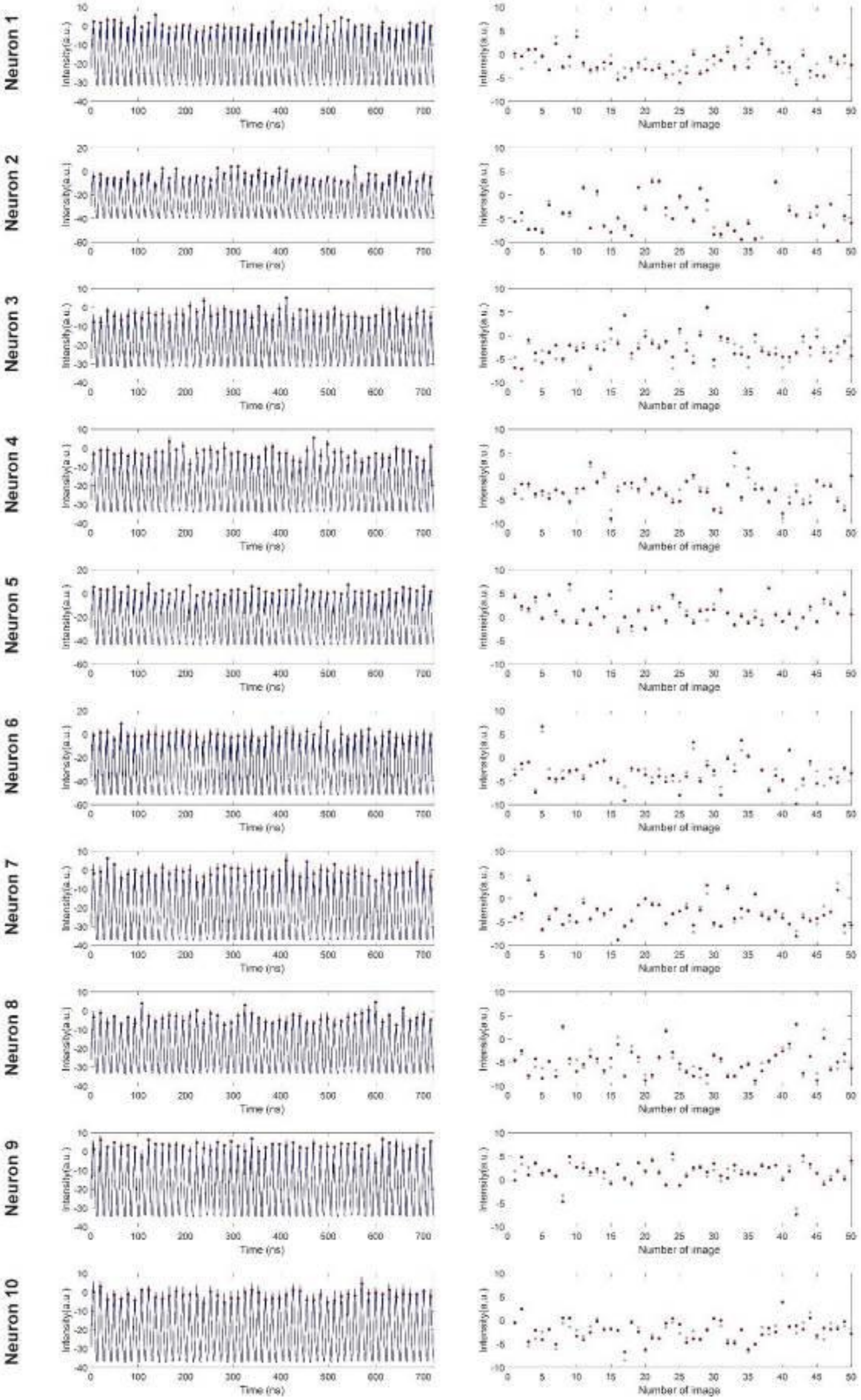

**Fig. 21.** Experimental results of the ten output neurons in the fully connected layer. The left graphs show the output waveforms, while the right graphs show the corresponding sampled (red dot) and theoretically calculated (grey dot) intensities of the neurons.

| Parameter / Approach | Input data dimension | Computing speed | | Scalability & reconfigurability* | Integrated components |
|---|---|---|---|---|---|
| | | FLOPS | Bit resolution# | | |
| **Lens [11]** | 784 | <1 | 8-bit | Level 1 | None |
| **Coupler array [8]** | 4 | <1 | N.A. | Level 2 | Weight and sum circuits |
| **Reservoir [9]** | 371 (up to 1113) | ~17.6 G | N.A. | Level 2 | None |
| **Phase change material [12]** | 15 | <1 | 1-bit | Level 2 | Weight and sum circuits |
| **Broadcast and weight [27]** | 8 | ~1 G | N.A. | Level 3 | None |
| **Time and wavelength interleaving (this work)** | 250,000 (theoretically unlimited) | 11.321 T | 8-bit | Level 4 | 90-wavelength light source |

**Bit resolution#**: this parameter, adopted from digital protocols, can roughly reflects the intensity resolution of analog ONNs. For example, analog ONNs with bit resolution = 8-bit can process input data with $2^8$=256 discrete levels, such as the benchmark MINIST handwritten digits. This parameter here for our work and other existing ONN is determined by the bit-resolution of the input data, and for some work is not available (N. A.) since the relevant details of the tested dataset is not provided.

**Scalability & reconfigurability***:

Level 1—can hardly reconfigure the synaptic weights; Level 2—the synaptic weights can be reconfigured, but the network structure, including the number of layers and neurons in each layer, can hardly be reconfigured; Level 3—the weights and network structure can be reconfigured, but the computing function of each layer cannot be reconfigured. Level 4—the weights and network structure can be dynamically reconfigured, the function of each layer can also be reconfigured (such as from convolutional layers to fully connected layers).

## Performance comparison

Here, we discuss the recent progress of optical neuromorphic hardware (Table 1). This section is not comprehensive but focuses on the leading results that address the most crucial technical issues for optical computing hardware. The input data dimension directly determines the complexity of the processing task. In real-life scenarios, the input data dimension is generally very large, for example, a human face image would require over 60,000 pixels. Thus, to make optical computing hardware eventually useful, the input data dimension would need to be at least over 20,000. In this work we demonstrate processing of images containing 250,000 pixels, which is 224 x higher than previous reports.

The computing speed is perhaps the most important parameter for computing hardware and is the main strength of optical approaches. Although there has not been a widely accepted definition of optical hardware computing speed, the key issue is the number of data sets that are processed within a certain time period - i.e., how many images can be processed per second. As such, although in some approaches [8, 11, 12], the latency is low due to the short physical path lengths, the computing speed remains very low due to the absence of high-speed data interfaces (i.e., input and output nodes are not updated at a high rate). Although other approaches [9, 27] offer high-speed data interfaces, their computing parallelism is not high and so their speed is similar to the input data rate. In our work, through the use of high-speed data interfaces (62.9 Giga Baud) and time-wavelength interleaving, we achieved a record computing speed of 11.321 Tera-OPS, > 500 x higher than previous reports.

Finally, the scalability and reconfigurability determines the versatility of the optical computing hardware. Approaches that cannot dynamically reconfigure the synapses [11] (marked as “Level 1” in the table) are barely trainable. Approaches at Level 2 [9, 12, 27] support online training, however, they can only process a specific task since the network structure is fixed once the device is fabricated. For approaches [27] at Level 3, different tasks can be processed although the function of each layer is fixed, which limits the hardware from implementing more complex operations other than matrix multiplication. Our work represents the first approach that operates at Level 4 with full dynamic reconfigurability in all respects. Here, the synaptic weights can be reconfigured by programming the WaveShaper.

Further, the number of synapses per neuron can be reconfigured by reallocating the wavelength channels with the de-multiplexer. The number of layers can be reconfigured by changing the number of stacked devices. Finally, the computing function can be switched between convolution and matrix multiplication by changing the sampling method. The degree of integration directly determines the potential computing density (processing capability per unit footprint). For approaches not well suited to integration [8, 11, 27], the potential computing density is low. While other approaches achieve limited integration of the weight and sum circuits [8, 12] - probably the most challenging issue — advanced integrated light sources have not been demonstrated. The performance of the light source directly determines the performance of the overall hardware in both input data scale [8] and number of synaptic connections per neuron [12]. The $mm^2$ sized microcomb offers a large number of precisely-spaced wavelengths, which enhances the overall parallelism and computing density, representing a major step towards the full integration of optical computing hardware.

## 4. DISCUSSION

This approach can be readily scaled in performance in terms of input data size, as well as network size and speed. The data size is limited in practice only by the memory of the electrical digital-to-analog converters, and so in principle it is possible to process 4K-resolution (4096×2160) images. By integrating 100 photonic convolution accelerators layers (still much less than the 65536 processors integrated in the Google TPU [21]), the optical CNN would be capable of solving much more difficult image recognition tasks at a vector computing speed of 100 × 11.3=1.130 Peta-OPS. Further, the optical CNN presented here supports online training, since the optical spectral shaper used to establish the synapses can be dynamically reconfigured as fast as 500 ms or faster with integrated optical spectral shapers [91].

Although we had a non-trivial optical latency of 0.11 μs introduced by dispersive fibre spool, this did not affect the operational speed. Moreover, the latency of the delay function can be virtually eliminated (to < 200 ps) by using integrated highly dispersive devices such as photonic crystals or customized chirped Bragg gratings [92] or even tunable dispersion compensators [93-96]. Finally, current nanofabrication techniques can enable significantly higher levels of integration of the convolutional accelerator. The micro-comb source itself is based on a CMOS compatible platform that is intrinsically designed for large-scale integration. Other components such as the optical spectral shaper, modulator, dispersive media, de-multiplexer and photodetector have all been realized in integrated form [91, 92, 98]. These devices will have a significant impact on the progress towards monolithic integration of RF and microwave systems, particularly those focused on optical Kerr microcombs [99-110].

## 5. CONCLUSION

We demonstrate a universal optical vector convolutional accelerator operating beyond 10 TOPS, generating convolutions of images of 250,000 pixels with 8-bit resolution for 10 kernels simultaneously — enough for facial image recognition. We then use the same hardware to sequentially form a deep optical CNN with ten output neurons, achieving successful recognition of full 10 digits with 900 pixel handwritten digit images with 88% accuracy. Our approach is scalable and trainable to more complex networks for demanding applications to unmanned vehicles and real-time video recognition.

Competing interests: The authors declare no competing interests.

## References

1. LeCun, Y., Bengio, Y. & Hinton, G. Deep learning. Nature **521**, 436–444 (2015).
2. Schalkoff, R. J. Pattern recognition. Wiley Encyclopedia of Computer Science and Engineering (2007).
3. Mnih, V. et al. Human-level control through deep reinforcement learning. Nature **518**, 529–533 (2015).
4. Silver, D. et al. Mastering the game of Go without human knowledge. Nature **550**, 354 (2017).
5. Krizhevsky, A., Sutskever, I. & Hinton, G. E. ImageNet Classification with Deep Convolutional Neural Networks. Commun Acm 60, 84-90 (2017).
6. Yao, P. et al. Fully hardware-implemented memristor convolutional neural network. Nature **577**, 641–646 (2020).
7. Lawrence, S. et al., “Face recognition: A convolutional neural-network approach”, IEEE transactions on neural networks 8, 98-113 (1997).
8. Shen, Y. et al. Deep learning with coherent nanophotonic circuits. Nature Photonics **11**, 441 (2018).
9. Larger, L. et al. High-speed photonic reservoir computing using a time-delay-based architecture: Million words per second classification. Phys. Rev. X **7**, 011015 (2017).
10. Peng, H., Nahmias, M. A., Lima, T. F. d., Tait, A. N. & Shastri, B. J. Neuromorphic Photonic Integrated Circuits. IEEE Journal of Selected Topics in Quantum Electronics **24**, 1-15, doi:10.1109/JSTQE.2018.2840448 (2018).
11. Lin, X. et al. All-optical machine learning using diffractive deep neural networks. Science **361**, 1004-1008 (2018).
12. Feldmann, J. et al., “All-optical spiking neurosynaptic networks with self-learning capabilities”, Nature **569**, 208-214 (2019).
13. X. Xu et al, “Photonic perceptron based on a Kerr microcomb for scalable high speed optical neural networks”, Laser and Photonics Reviews, vol. 14, no. 8, 2000070 (2020). DOI: 10.1002/lpor.202000070.
14. X. Xu, et al., “11 TOPs photonic convolutional accelerator for optical neural networks”, Nature **589**, 44-51 (2021).

15. Feldmann, J. *et al.*, "Parallel convolutional processing using an integrated photonic tensor core", Nature **589**, 52-58 (2021).
16. B. J. Shastri *et al.,* "Photonics for artificial intelligence and neuromorphic computing", Nature Photonics **15,** (2) 102-114 (2021).
17. H. Wu, and Q. Dai, "Artificial intelligence accelerated by light", Nature **589**, 25-26 (2021).
18. G. Wetzstein *et al.,* "Inference in artificial intelligence with deep optics and photonics", Nature **588** (7836), 39-47 (2020).
19. Ambrogio, S. *et al.,* "Equivalent-accuracy accelerated neural-network training using analogue memory", *Nature* **558**, 60 (2018).
20. Esser, S. K. et al., "Convolutional networks for fast, energy-efficient neuromorphic computing", Proc. of the National Academy of Sciences **113**, 11441 (2016).
21. Graves, A. et al. Hybrid computing using a neural network with dynamic external memory. Nature **538**, 471–476 (2016).
22. Miller, D. A. B. Attojoule Optoelectronics for Low-Energy Information Processing and Communications. Journal of Lightwave Technology **35**, 346-396 (2017).
23. Appeltant, L. et al. Information processing using a single dynamical node as complex system. Nature Communications **2**, 468 (2011).
24. Chang, J., Sitzmann, V., Dun, X., Heidrich, W. & Wetzstein, G. Hybrid optical-electronic convolutional neural networks with optimized diffractive optics for image classification. Scientific Reports 8 (2018).
25. Vandoorne, K. et al., "Experimental demonstration of reservoir computing on a silicon photonics chip", Nature Communications **5**, 3541 (2014).
26. Brunner, D. et al., "Parallel photonic information processing at gigabyte per second data rates using transient states", Nature Communications **4**, 1364 (2013).
27. Tait, A. N. et al., "Demonstration of WDM weighted addition for principal component analysis", Optics Express **23**, 12758-12765 (2015).
28. Pasquazi, A. et al. Micro-combs: a novel generation of optical sources. Physics Reports **729**, 1-81 (2018).
29. Moss, D. J. et al., "New CMOS-compatible platforms based on silicon nitride and Hydex for nonlinear optics", Nature photonics **7**, 597 (2013).
30. Kippenberg, T. J., Gaeta, A. L., Lipson, M. & Gorodetsky, M. L. Dissipative Kerr solitons in optical microresonators. Science **361**, 567 (2018).
31. Savchenkov, A. A. et al. Tunable optical frequency comb with a crystalline whispering gallery mode resonator. Physics Review Letters **101**, 093902 (2008).
32. Spencer, D. T. et al. An optical-frequency synthesizer using integrated photonics. Nature **557**, 81-85 (2018).
33. Marin-Palomo, P. et al. Microresonator-based solitons for massively parallel coherent optical communications. Nature **546**, 274 (2017).
34. B. Corcoran, et al., "Ultra-dense optical data transmission over standard fiber with a single chip source", Nature Communications, vol. 11, Article:2568, 2020.
35. Kues, M. et al. Quantum optical microcombs. Nature Photonics **13**, (3) 170-179 (2019). doi:10.1038/s41566-019-0363-0
36. C. Reimer, L. Caspani, M. Clerici, et al., "Integrated frequency comb source of heralded single photons," Optics Express, vol. 22, no. 6, pp. 6535-6546, 2014.
37. C. Reimer, et al., "Cross-polarized photon-pair generation and bi-chromatically pumped optical parametric oscillation on a chip", Nature Communications, vol. 6, Article 8236, 2015. DOI: 10.1038/ncomms9236.
38. L. Caspani, C. Reimer, M. Kues, et al., "Multifrequency sources of quantum correlated photon pairs on-chip: a path toward integrated Quantum Frequency Combs," Nanophotonics, vol. 5, no. 2, pp. 351-362, 2016.
39. C. Reimer et al., "Generation of multiphoton entangled quantum states by means of integrated frequency combs," Science, vol. 351, no. 6278, pp. 1176-1180, 2016.
40. M. Kues, et al., "On-chip generation of high-dimensional entangled quantum states and their coherent control", Nature, vol. 546, no. 7660, pp. 622-626, 2017.
41. P. Roztocki et al., "Practical system for the generation of pulsed quantum frequency combs," Optics Express, vol. 25, no. 16, pp. 18940-18949, 2017.
42. Y. Zhang, et al., "Induced photon correlations through superposition of two four-wave mixing processes in integrated cavities", Laser and Photonics Reviews, vol. 14, no. 7, pp. 2000128, 2020. DOI: 10.1002/lpor.202000128
43. C. Reimer, et al.,"High-dimensional one-way quantum processing implemented on d-level cluster states", Nature Physics, vol. 15, no.2, pp. 148–153, 2019.
44. Stern, B., Ji, X., Okawachi, Y., Gaeta, A. L. & Lipson, M. Battery-operated integrated frequency comb generator. Nature **562**, 401 (2018).
45. H. Bao, et al., Laser cavity-soliton microcombs, Nature Photonics, vol. 13, no. 6, pp. 384-389, Jun. 2019.
46. Lugiato, L. A., Prati, F. & Brambilla, M. Nonlinear Optical Systems, (Cambridge University Press, 2015).
47. Cole, D. C., Lamb, E. S., Del'Haye, P., Diddams, S. A. & Papp, S. B. Soliton crystals in Kerr resonators. Nature Photonics 11, 671 (2017).
48. Wang, W., et al.., Robust soliton crystals in a thermally controlled microresonator, Opt. Lett., 43, 2002 (2018).
49. T. Monro, D.J.Moss, M. Bazylenko, C. Martijn de Sterke, and L. Poladian, "Observation of self-trapping of light in a self written channel in photosensitive glass", Physical Review Letters, vol. 80, 4072 (1998).
50. M. Ferrera et al., "CMOS compatible integrated all-optical RF spectrum analyzer", Optics Express, vol. 22, no. 18, 21488 - 21498 (2014).
51. A. Pasquazi, et al., "Sub-picosecond phase-sensitive optical pulse characterization on a chip", Nature Photonics, vol. 5, no. 10, pp. 618-623 (2011).
52. M. Kues, et al., "Passively modelocked laser with an ultra-narrow spectral width", Nature Photonics, vol. 11, no. 3, pp. 159, 2017.
53. L. Razzari, et al., "CMOS-compatible integrated optical hyper-parametric oscillator," Nature Photonics, vol. 4, no. 1, pp. 41-45, 2010.
54. M. Ferrera, et al., "Low-power continuous-wave nonlinear optics in doped silica glass integrated waveguide structures," Nature Photonics, vol. 2, no. 12, pp. 737-740, 2008.

55. M.Ferrera et al."On-Chip ultra-fast 1st and 2nd order CMOS compatible all-optical integration", Opt. Express, vol. 19, (23)pp. 23153-23161 (2011).
56. D. Duchesne, M. Peccianti, M. R. E. Lamont, et al., "Supercontinuum generation in a high index doped silica glass spiral waveguide," Optics Express, vol. 18, no, 2, pp. 923-930, 2010.
57. H Bao, L Olivieri, M Rowley, ST Chu, BE Little, R Morandotti, DJ Moss, ... "Turing patterns in a fiber laser with a nested microresonator: Robust and controllable microcomb generation", Physical Review Research **2** (2), 023395 (2020).
58. M. Ferrera, et al., "On-chip CMOS-compatible all-optical integrator", Nature Communications, vol. 1, Article 29, 2010.
59. A. Pasquazi, et al., "All-optical wavelength conversion in an integrated ring resonator," Optics Express, vol. 18, no. 4, pp. 3858-3863, 2010.
60. A. Pasquazi, Y. Park, J. Azana, et al., "Efficient wavelength conversion and net parametric gain via Four Wave Mixing in a high index doped silica waveguide," Optics Express, vol. 18, no. 8, pp. 7634-7641, 2010.
61. M. Peccianti, M. Ferrera, L. Razzari, et al., "Subpicosecond optical pulse compression via an integrated nonlinear chirper," Optics Express, vol. 18, no. 8, pp. 7625-7633, 2010.
62. Little, B. E. et al., "Very high-order microring resonator filters for WDM applications", IEEE Photonics Technol. Lett. **16**, 2263–2265 (2004).
63. M. Ferrera et al., "Low Power CW Parametric Mixing in a Low Dispersion High Index Doped Silica Glass Micro-Ring Resonator with Q-factor > 1 Million", Optics Express, vol.17, no. 16, pp. 14098–14103 (2009).
64. M. Peccianti, et al., "Demonstration of an ultrafast nonlinear microcavity modelocked laser", Nature Communications, vol. 3, pp. 765, 2012.
65. A. Pasquazi, et al., "Self-locked optical parametric oscillation in a CMOS compatible microring resonator: a route to robust optical frequency comb generation on a chip," Optics Express, vol. 21, no. 11, pp. 13333-13341, 2013.
66. A. Pasquazi, et al., "Stable, dual mode, high repetition rate mode-locked laser based on a microring resonator," Optics Express, vol. 20, no. 24, pp. 27355-27362, 2012.
67. Wu, J. *et al.* RF Photonics: An Optical Microcombs' Perspective. IEEE Journal of Selected Topics in Quantum Electronics **24**, 1-20 (2018).
68. Xu, X., et al., Photonic microwave true time delays for phased array antennas using a 49 GHz FSR integrated micro-comb source, *Photonics Research*, **6**, B30-B36 (2018).
69. T. G. Nguyen *et al.*, "Integrated frequency comb source-based Hilbert transformer for wideband microwave photonic phase analysis," *Opt. Express*, vol. 23, no. 17, pp. 22087-22097, Aug. 2015.
70. X. Xu, J. Wu, M. Shoeiby, T. G. Nguyen, S. T. Chu, B. E. Little, R. Morandotti, A. Mitchell, and D. J. Moss, "Reconfigurable broadband microwave photonic intensity differentiator based on an integrated optical frequency comb source," *APL Photonics*, vol. 2, no. 9, 096104, Sep. 2017.
71. X. Xu, M. Tan, J. Wu, R. Morandotti, A. Mitchell, and D. J. Moss, "Microcomb-based photonic RF signal processing", *IEEE Photonics Technology Letters*, vol. 31 no. 23 1854-1857, 2019.
72. X. Xu, *et al.*, "Broadband RF channelizer based on an integrated optical frequency Kerr comb source," *Journal of Lightwave Technology,* vol. 36, no. 19, pp. 4519-4526, 2018.
73. X. Xu, *et al.*, "Continuously tunable orthogonally polarized RF optical single sideband generator based on micro-ring resonators," *Journal of Optics,* vol. 20, no. 11, 115701. 2018.
74. X. Xu, *et al.*, "Orthogonally polarized RF optical single sideband generation and dual-channel equalization based on an integrated microring resonator," *Journal of Lightwave Technology,* vol. 36, no. 20, pp. 4808-4818. 2018.
75. X. Xu, *et al.*, "Photonic microwave true time delays for phased array antennas using a 49 GHz FSR integrated optical micro-comb source," *Photonics Res,* vol. 6, no. 5, pp. B30-B36, 2018.
76. X. Xu, *et al.*, "Advanced adaptive photonic RF filters with 80 taps based on an integrated optical micro-comb source," *Journal of Lightwave Technology,* vol. 37, no. 4, pp. 1288-1295, 2019.
77. X. Xu, *et al.,* Broadband microwave frequency conversion based on an integrated optical micro-comb source", *Journal of Lightwave Technology,* vol. 38 no. 2, pp. 332-338, 2020.
78. M. Tan, *et al.,* "Photonic RF and microwave filters based on 49GHz and 200GHz Kerr microcombs", *Optics Comm*. vol. 465,125563, Feb. 22. 2020.
79. X. Xu, *et al.,* "Broadband photonic RF channelizer with 90 channels based on a soliton crystal microcomb", *Journal of Lightwave Technology,* Vol. 38, no. 18, pp. 5116 - 5121, 2020. doi: 10.1109/JLT.2020.2997699.
80. X. Xu, *et al.*, "Photonic RF and microwave integrator with soliton crystal microcombs", *IEEE Transactions on Circuits and Systems II: Express Briefs,* vol. 67, no. 12, pp. 3582-3586, 2020. DOI:10.1109/TCSII.2020.2995682.
81. X. Xu, *et al.,* "Photonic RF phase-encoded signal generation with a microcomb source", *J. Lightwave Technology,* vol. 38, no. 7, 1722-1727, 2020.
82. X. Xu, *et al.*, "High performance RF filters via bandwidth scaling with Kerr micro-combs," *APL Photonics,* vol. 4, no. 2, pp. 026102. 2019.
83. M. Tan, *et al.,* "Microwave and RF photonic fractional Hilbert transformer based on a 50 GHz Kerr micro-comb", *Journal of Lightwave Technology,* vol. 37, no. 24, pp. 6097 – 6104, 2019.
84. M. Tan, *et al.,* "RF and microwave fractional differentiator based on photonics", *IEEE Transactions on Circuits and Systems: Express Briefs,* vol. 67, no.11, pp. 2767-2771, 2020. DOI:10.1109/TCSII.2020.2965158.

85. M. Tan, *et al.*, "Photonic RF arbitrary waveform generator based on a soliton crystal micro-comb source", Journal of Lightwave Technology, vol. 38, no. 22, pp. 6221-6226, Oct 22. 2020. DOI: 10.1109/JLT.2020.3009655.
86. M. Tan, X. Xu, J. Wu, R. Morandotti, A. Mitchell, and D. J. Moss, "RF and microwave high bandwidth signal processing based on Kerr Micro-combs", Advances in Physics X, VOL. 6, NO. 1, 1838946 (2021). DOI:10.1080/23746149.2020.1838946.
87. X. Xu, et al., "Advanced RF and microwave functions based on an integrated optical frequency comb source," Opt. Express, vol. 26 (3) 2569 2018.
88. Krizhevsky A, Sutskever I, Hinton G E. Imagenet classification with deep convolutional neural networks. Advances in neural information processing systems **25**, 1097-1105 (2012).
89. LeCun, Y., Bottou, L., Bengio, Y. & Haffner, P. Gradient-based learning applied to document recognition. Proceedings of the IEEE **86**, 2278-2324 (1998).
90. Bishop, C. M. Neural networks for pattern recognition. (Oxford university press, 1995).
91. Metcalf, A. J. et al. Integrated line-by-line optical pulse shaper for high-fidelity and rapidly reconfigurable RF-filtering. Optics Express **24**, 23925-23940 (2016).
92. Sahin, E et al., "Large, scalable dispersion engineering using cladding-modulated Bragg gratings on a silicon chip", Applied Physics Letters **110**, 161113 (2017).
93. D. J. Moss et al., "Tunable dispersion and dispersion slope compensators for 10Gb/s using all-pass multicavity etalons", IEEE Phot. Technology Letters, vol. 15, no. 5, 730-732 (2003).
94. L.M. Lunardi et al., "Tunable dispersion compensators based on multi-cavity all-pass etalons for 40Gb/s systems", J. Lightwave Technology, vol. 20, (12) 2136 (2002).
95. D. J. Moss, et al., "Multichannel tunable dispersion compensation using all-pass multicavity etalons", paper TuT2 Optical Fiber Communications Conference, Anaheim (2002). Post-conference Technical Digest (IEEE Cat. No.02CH37339). Opt Soc. America. Part vol.1, 2002, pp. 132-3. Washington, DC, USA.
96. L.M. Lunardi, D. Moss, S. Chandrasekhar, L.L. Buhl, "An etalon based tunable dispersion compensator (TDC) device for 40Gb/s applications", European Conference on Optical Communications (ECOC), paper 5.4.6, Copenhagen, Sept. (2002). IEEE. Part vol. 2, 2002, pp. 2 vol. 2. Piscataway, NJ, USA. INSPEC Accession Number: 9153476, Print ISBN: 87-90974-63-8.
97. Wang, C. et al. Integrated lithium niobate electro-optic modulators operating at CMOS-compatible voltages. Nature **562**, 101 (2018).
98. T.Ido, H.Sano, D.J.Moss, S.Tanaka, and A.Takai, "Strained InGaAs/InAlAs MQW electroabsorption modulators with large bandwidth and low driving voltage", IEEE Photonics Technology Letters, vol. 6, 1207 (1994). DOI: 10.1109/68.329640.
99. Mengxi Tan, X. Xu, J. Wu, T. G. Nguyen, S. T. Chu, B. E. Little, R. Morandotti, A. Mitchell, and David J. Moss, "Photonic Radio Frequency Channelizers based on Kerr Optical Micro-combs", Journal of Semiconductors, vol. 42, no. 4, 041302 (2021). (ISSN 1674-4926). DOI:10.1088/1674-4926/42/4/041302.
100. H.Bao, L.Olivieri, M.Rowley, S.T. Chu, B.E. Little, R.Morandotti, D.J. Moss, J.S.T. Gongora, M.Peccianti and A.Pasquazi, "Laser Cavity Solitons and Turing Patterns in Microresonator Filtered Lasers: properties and perspectives", Paper No. LA203-5, Paper No. 11672-5, SPIE LASE, SPIE Photonics West, San Francisco CA March 6-11 (2021). DOI:10.1117/12.2576645
101. Mengxi Tan, X. Xu, J. Wu, A. Boes, T. G. Nguyen, S. T. Chu, B. E. Little, R. Morandotti, A. Mitchell, and David J. Moss, "Advanced microwave signal generation and processing with soliton crystal microcombs", or "Photonic convolutional accelerator and neural network in the Tera-OPs regime based on Kerr microcombs", Paper No. 11689-38, PW21O-OE201-67, Integrated Optics: Devices, Materials, and Technologies XXV, SPIE Photonics West, San Francisco CA March 6-11 (2021). DOI: 10.1117/12.2584017
102. Mengxi Tan, Bill Corcoran, Xingyuan Xu, Andrew Boes, Jiayang Wu, Thach Nguyen, Sai T. Chu, Brent E. Little, Roberto Morandotti, Arnan Mitchell, and David J. Moss, "Optical data transmission at 40 Terabits/s with a Kerr soliton crystal microcomb", Paper No.11713-8, PW21O-OE803-23, Next-Generation Optical Communication: Components, Sub-Systems, and Systems X, SPIE Photonics West, San Francisco CA March 6-11 (2021). DOI:10.1117/12.2584014
103. Mengxi Tan, X. Xu, J. Wu, A. Boes, T. G. Nguyen, S. T. Chu, B. E. Little, R. Morandotti, A. Mitchell, and David J. Moss, "RF and microwave photonic, fractional differentiation, integration, and Hilbert transforms based on Kerr micro-combs", Paper No. 11713-16, PW21O-OE803-24, Next-Generation Optical Communication: Components, Sub-Systems, and Systems X, SPIE Photonics West, San Francisco CA March 6-11 (2021). DOI:10.1117/12.2584018
104. Mengxi Tan, X. Xu, J. Wu, A. Boes, T. G. Nguyen, S. T. Chu, B. E. Little, R. Morandotti, A. Mitchell, and David J. Moss, "Broadband photonic RF channelizer with 90 channels based on a soliton crystal microcomb", or "Photonic microwave and RF channelizers based on Kerr micro-combs", Paper No. 11685-22, PW21O-OE106-49, Terahertz, RF, Millimeter, and Submillimeter-Wave Technology and Applications XIV, SPIE Photonics West, San Francisco CA March 6-11 (2021). DOI:10.1117/12.2584015
105. X. Xu, M. Tan, J. Wu, S. T. Chu, B. E. Little, R. Morandotti, A. Mitchell, B. Corcoran, D. Hicks, and D. J. Moss, "Photonic perceptron based on a Kerr microcomb for scalable high speed optical neural networks", IEEE Topical Meeting on Microwave Photonics (MPW), pp. 220-224,.Matsue, Japan, November 24-26, 2020. Electronic ISBN:978-4-88552-331-1. DOI: 10.23919/MWP48676.2020.9314409
106. Mengxi Tan, Bill Corcoran, Xingyuan Xu, Andrew Boes, Jiayang Wu, Thach Nguyen, S.T. Chu, B. E. Little, Roberto Morandotti, Arnan Mitchell, and David J. Moss, "Ultra-high bandwidth optical data transmission with a microcomb", IEEE Topical Meeting on Microwave Photonics (MPW), pp. 78-82.Virtual Conf., Matsue, Japan, November 24-26, 2020. Electronic ISBN:978-4-88552-331-1. DOI: 10.23919/MWP48676.2020.9314476
107. M. Tan, X. Xu, J. Wu, R. Morandotti, A. Mitchell, and D. J. Moss, "RF and microwave high bandwidth signal processing based on Kerr Micro-combs", Advances in Physics X, VOL. 6, NO. 1, 1838946 (2020). DOI:10.1080/23746149.2020.1838946.

108. Mengxi Tan, Xingyuan Xu, Jiayang Wu, Thach G. Nguyen, Sai T. Chu, Brent E. Little, Roberto Morandotti, Arnan Mitchell, and David J. Moss, "Photonic Radio Frequency Channelizers based on Kerr Micro-combs and Integrated Micro-ring Resonators", JOSarXiv.202010.0002.
109. Mengxi Tan, Xingyuan Xu, David Moss "Tunable Broadband RF Photonic Fractional Hilbert Transformer Based on a Soliton Crystal Microcomb", Preprints, DOI: 10.20944/preprints202104.0162.v1
110. Mengxi Tan, X. Xu, J. Wu, T. G. Nguyen, S. T. Chu, B. E. Little, R. Morandotti, A. Mitchell, and David J. Moss, "Orthogonally polarized Photonic Radio Frequency single sideband generation with integrated micro-ring resonators", Journal of Semiconductors, vol. 42 (4), 041305 (2021). DOI: 10.1088/1674-4926/42/4/041305.